\documentclass[onecolumn,floatfix,superscriptaddress,a4paper,
               showpacs,showkeys,nofootinbib,preprint]{revtex4}
\usepackage{epsfig}
\usepackage{latexsym}
\usepackage{xspace}
\usepackage{hyperref}
\usepackage[latin2]{inputenc}
\usepackage{indentfirst}
\usepackage{enumerate}

\usepackage{amsmath}
\usepackage{amssymb}
\usepackage[english]{babel}
\usepackage{url}

\graphicspath{{./graphics/}}

\newcommand{\eq}[1]{\begin{align} #1 \end{align}}

\begin{document}
\title{Cumulative production of pions by heavy baryonic resonances\\ in proton-nucleus collisions
}

\author{A.~Motornenko}
\affiliation{
Taras Shevchenko National University of Kiev, Kiev, Ukraine}

\author{M. I. Gorenstein}
 \affiliation{
Bogolyubov Institute for Theoretical Physics, Kiev, Ukraine}
\affiliation{
Frankfurt Institute for Advanced Studies, Johann Wolfgang Goethe University, Frankfurt, Germany}

\begin{abstract}
Pion production in proton-nucleus (p+A) collisions  
outside the kinematical boundary of proton-nucleon (p+N) reactions,
the so-called cumulative effect, is studied. Restrictions 
from energy-momentum conservation on the energy
of pions emitted in the backward direction in the target rest frame are
analyzed. It is assumed that the cumulative pions are produced 
in p+A reactions by heavy baryonic resonances.
The baryonic resonances are first created in p+N reactions.  Due to successive
collisions with nuclear nucleons the masses of these resonances may then 
increase and, simultaneously, their longitudinal velocities decrease. 
We also use the Ultra relativistic Quantum Molecular
Dynamics model to reveal the key role
of successive collisions of baryonic resonances with nuclear nucleons
for cumulative pion production in p+A reactions.
Further experimental studies of cumulative hadron production in p+A reactions at
high collision energies are needed to search for heavy hadron-like objects
and investigate their properties.

\end{abstract}



\maketitle

\section{Introduction}
About 40 years ago the so-called cumulative effect
was discovered in relativistic proton-nucleus
(p+A) collisions, i.e., secondary particles were
detected in the kinematic region
forbidden within an interaction of the projectile
proton with a free nucleon
\cite{baldin-74,baldin-77,leksin-77, frankel-79, bayukov-79}. Different theoretical models
were proposed  \cite{burov-77, gor-77, voloshyn, frankel, frankfurt, luk-79, efremov, efremov-88, QCD, gor, gor-80, gor-82, shmonin, kalinkin, re-1, re-2, re-3, bond}, however, a physical origin of this
effect is still not settled.
In the present paper the cumulative effect is considered
for pion production in inclusive reactions
p+A$\rightarrow \pi(180^\circ)+X$ with the pion emitted at $180^{\circ}$,
i.e., in the backward direction
in the rest frame of nuclei A.  First experiments on cumulative particle production
\cite{baldin-74, baldin-77, leksin-77} were performed at Synchrophasotron accelerator
of the Joint Institute for Nuclear Research in Dubna. The momenta of projectile protons
were taken as $p_0=6$~GeV/c and $8.4$~GeV/c. The p+A reactions with $p_0=400$~GeV/c
were then studied \cite{frankel-79, bayukov-79} at Fermilab, USA. In both experiments
atomic numbers of nuclear target were varied in a wide
region $A=10\div 200$.

Let $E_{\pi}^{*}$ denotes the maximal possible energy of the pion emitted
at angle $180^{\circ}$ in the laboratory frame
in a proton-nucleon (p+N) interaction at fixed projectile proton
momentum $p_0$. In p+A collisions at the same projectile proton momentum $p_0$,
pions emitted at  $180^{\circ}$
in the laboratory frame (this frame  coincides
with the nucleus rest frame) with energies $E_{\pi}> E_{\pi}^*$,
even above 2$E_{\pi}^*$, were experimentally observed \cite{baldin-74,baldin-77,leksin-77, frankel-79, bayukov-79}.

The results of Ref.~\cite{burov-77} demonstrated that Fermi motion of nuclear nucleons
can not describe the data of the cumulative pion production.
In Ref.~\cite{gor-77} a fireball model of cumulative effect was proposed.
According to that model, particle production in the cumulative
region $E_{\pi}> E_{\pi}^*$ is due to the formation of a massive hadronic fireball,
successive collisions of this fireball with nuclear nucleons, and
finally its  decay
with emission of a cumulative particle. In the present paper we
critically analyze the main assumptions of that approach.
We clarify the basic restrictions
on the energy $E_\pi$ for the backward
pion production in p+N and p+A reactions.
These are restrictions that follow from energy-momentum conservation.

Particle production in most phenomenological models,
e.g., in the popular relativistic transport models --
Ultra relativistic
Quantum Molecular Dynamics (UrQMD) \cite{urqmd}
and Hadron String Dynamics (HSD)
\cite{hsd} --  takes place exclusively
by means of binary reactions,
\eq{\label{reac}
h_1~+~h_2 ~\rightarrow ~ H_1~+~H_2~,
}
with subsequent decays of $H_1$ and/or $H_2$ excited states.
These reactions satisfy energy-momentum conservation,
as well as conservations of baryonic number, electric charge, and strangeness.
In elementary particle collisions, like p+N,
the  excited states $H_1$ and/or $H_2$ are hadronic resonances
and/or strings, while $h_1$ and $h_2$ are stable hadrons.
In p+A and A+A collisions, $h_1$ or/and $h_2$ can be also resonance(s),
and $H_1$ or/and $H_2$ can be the stable hadron(s). Therefore, any relations
between masses of $h_1$, $h_2$ and $H_1$, $H_2$ in Eq.~(\ref{reac}) are possible
in the course of p+A and A+A reactions.
Among secondary inelastic reactions there are also the following
\eq{\label{reac1}
h_1~+~h_2 ~\rightarrow ~ H~,
}
where $h_1$ and $h_2$ can be both mesons, baryon and antibaryon
($H$ is the meson resonance or mesonic string in these cases), or meson and baryon
($H$ is the baryonic resonance or baryonic string).
Experimental information on the quantum numbers and decay channels
of mesonic and baryonic resonances
is available, and the string  model \cite{string} is assumed to describe
the formation and decay of strings.

The main physical quantities in our analysis are the masses and longitudinal
(i.e. along the collision axis) velocities
of the baryonic resonances created in reactions~(\ref{reac}).
In p+A collisions, resonances $R$ are first produced in p+N$\rightarrow R$+N reactions,
and later they participate in successive $R$+N$\rightarrow R'$+N collisions.
The evolution of these  quantities due to subsequent collisions of the resonances
with nuclear nucleons  in the course of p+A reaction is investigated.
It is argued that the cumulative pions in p+A reactions are created
by baryonic resonances with very high masses that are formed due to successive collisions 
with nuclear nucleons.
We also use the UrQMD model to describe the existing data
and analyze some microscopic aspects of cumulative pion production in p+A reactions.

The paper is organized as follows.  In Sec.~\ref{sec-kinem} the basic
restrictions from relativistic kinematics on pion production
in p+N  reactions are considered.  In Sec.~\ref{sec-multi} the kinematic restrictions
on pion production in p+A collisions are discussed. The role of successive collisions
of a baryonic resonance created in p+N reaction with other nucleons inside
the nucleus target is investigated.  In Sec.~\ref{sec-UrQMD} the UrQMD model
is applied to backward pion
production in p+p and p+A reactions.
A summary in Sec.~\ref{sec-sum} closes the article.

\section{Kinematic restrictions in N+N reactions}\label{sec-kinem}
\begin{figure}[h!]
\centering
    \includegraphics[width=1\textwidth]{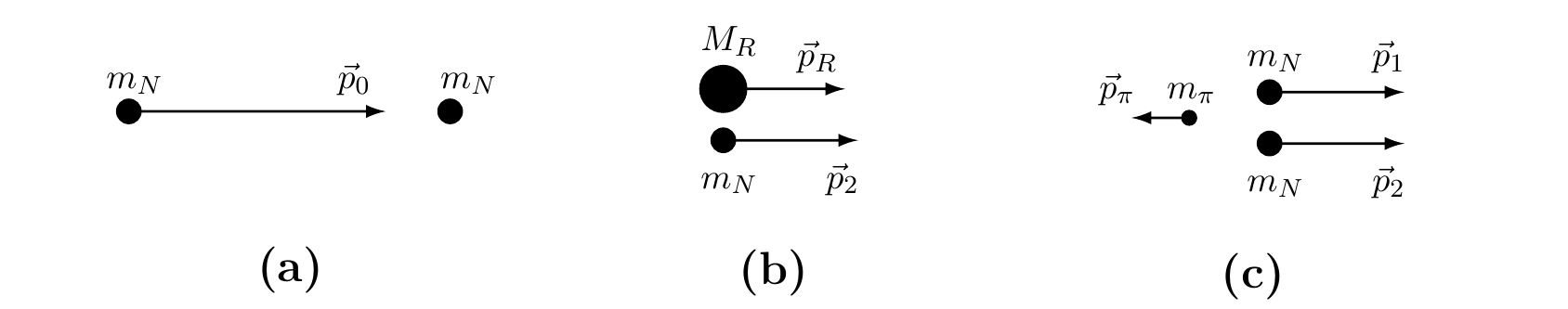}
    \caption{Pion production in NN$\rightarrow$NN$\pi(180^\circ)$ reaction.
    (a): initial stage.  (b): intermediate stage. (c): final stage.
      }
    \label{fig-pN}
\end{figure}

In this section we consider general
restrictions on energy $E_\pi$ of the pion emitted in the backward direction
(i.e., at $180^\circ$) in p+N reactions in the target rest frame.  These restrictions
are consequences of the energy-momentum conservation.
We are interested in
finding a maximal value of the pion energy.
Two nucleons should be anyhow  present in a final state because of baryonic number conservation.
It is evident that the production of any additional hadron(s) and/or a presence of non-zero
transverse momenta in the final state of two nucleons would require an additional energy
and lead to a reduction of $E_\pi$ value at fixed projectile momentum $p_0$.
Thus, to find the maximum of $E_{\pi}$ one should consider only the reaction
p+N$\rightarrow\pi(180^\circ)$+N+N and restrict the kinematical analysis to the one-dimensional
(longitudinal) direction, i.e., all particle momenta should be directed along the collision axis.
The emission of $\pi$-meson at $180^{\circ}$
is mediated by a baryonic resonance state.
Different stages of the NN$\rightarrow$NN$\pi(180^\circ)$ reaction are schematically
presented in Fig.~\ref{fig-pN}.

The energy-momentum conservation in the target rest frame reads\footnote{We use values t
he $m_N\cong 0.94$~GeV$\mathrm{/c^2}$
and $m_\pi \cong 0.14$~GeV$\mathrm{/c^2}$
for the nucleon and pion masses, respectively, neglecting  small mass differences for different
isospin states.}:
\eq{\label{pN-cons}
\sqrt{p_0^2+m_N^2}+m_N= \sqrt{p_\pi^2+m_\pi^2}+\sqrt{p_1^2+m_N^2}+\sqrt{p_2^2+m_N^2}~,~~~~~
 p_{0}=p_1+p_2-p_\pi~,
 %\label{p-cons}
%
}
where $p_1$ and $p_2$  are absolute values of
the final state  nucleon momenta, and $p_\pi=\sqrt{(E_\pi)^2-m_\pi^2}$ is that
of the pion emitted in the backward direction.
Substituting $p_2$ from the second equation of Eq.~(\ref{pN-cons}) to the first one
the value of $E_\pi$ is presented as a function of $p_0$ and $p_1$.
At fixed $p_0$, the pion energy $E_\pi$
reaches its maximal value denoted as $E_{\pi}^*$ at the point
where $\partial E_\pi/\partial p_1=0$. One finds
\eq{\label{p*1}
p^*_1 \equiv p_1=p_2= \frac{p_0+p^*_{\pi}}{2}~,
}
thus,
both nucleons should have the same values of final momenta.
The maximal pion energy is
then given by an implicit equation\footnote{Equation (\ref{E-max}) can be transformed
to a 4th order algebraic equation. Its explicit solution is, however, too cumbersome
and gives no advantages.}
\eq{\label{E-max}
E_{\pi}^*=m_N+\sqrt{m_N^2+p_0^2}- 2\,\sqrt{m_N^2+
\left(\frac{p_0+
\sqrt{(E_{\pi}^*)^2-m_{\pi}^2}}{2}\right)^2}~.}
%(p_1^*)^2}~.

\begin{figure}[h!]
\centering
    \includegraphics[width=0.65\textwidth]{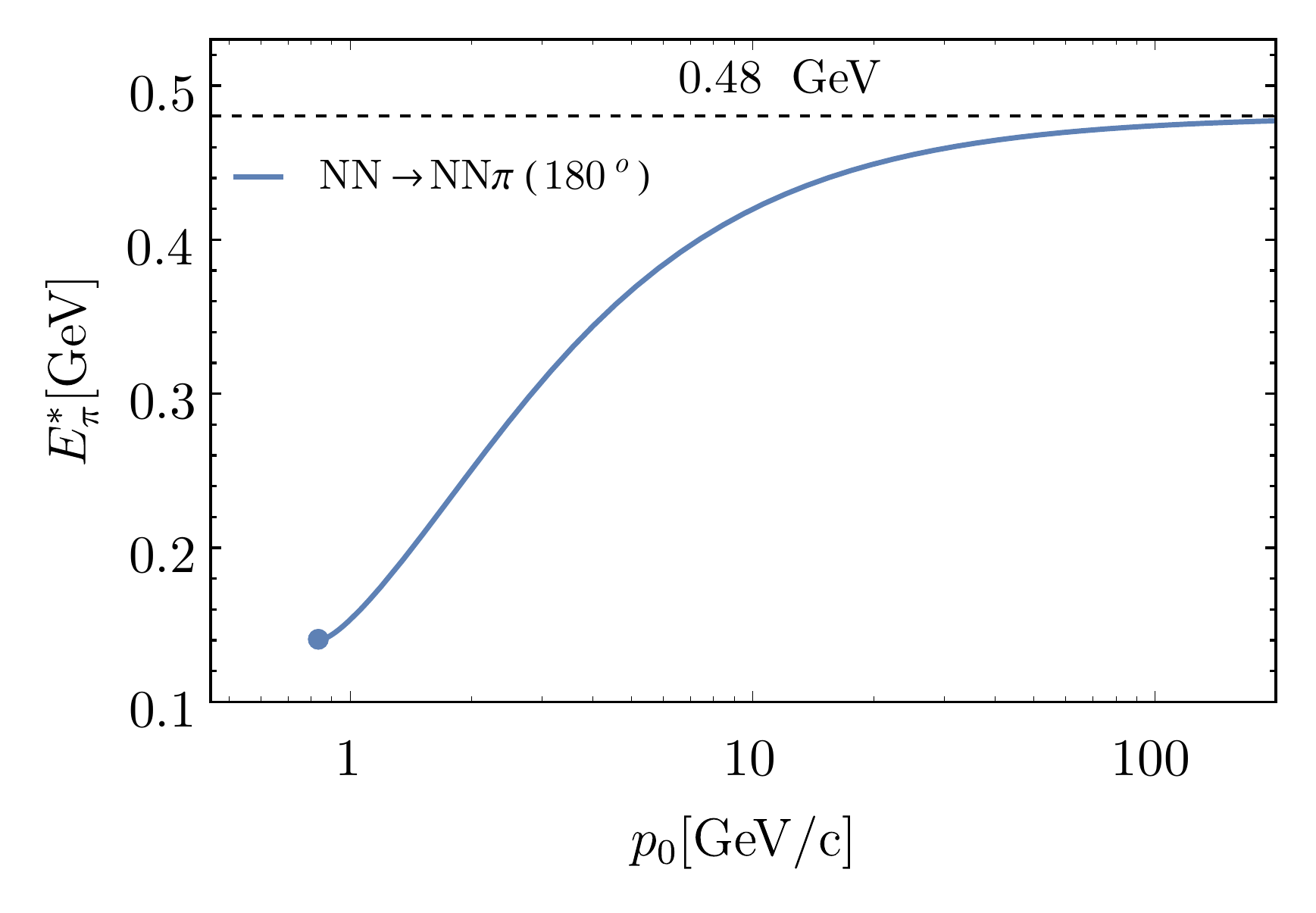}
    \caption{The maximal pion energy $E_\pi^*$ in
     NN$\rightarrow$NN$\pi(180^\circ)$ reaction presented in Fig.~\ref{fig-pN}.
    $E_\pi^*$ is calculated from Eq.~(\ref{E-max}) as a function of projectile proton momentum $p_0$.
    A threshold value $p_{0}\cong 0.83$~GeV/c
    corresponds to $p_\pi=0$. Horizontal dashed line presents the upper limit for $E_\pi^*$ at
    $p_0\rightarrow \infty$. 
    }
    \label{fig-NN}
\end{figure}

%
The index 1 in  $p_1^*$ (also in $M_1^*$ and $v^*_1$ below)
reminds that these quantities are obtained in a collision of
the projectile proton with {\it one} nucleon.
The solution of Eq.~(\ref{E-max}) for $E^*_\pi$  is presented in Fig.~\ref{fig-NN} as a function of $p_0$.

A minimal (threshold) value of  $p_0\cong 0.83$~GeV/c is
needed to create a pion with $p_\pi=0$.
For $p_0>0.83$~GeV/c the backward pion production with non-zero momenta $p_\pi$
becomes possible.
As seen from Fig.~\ref{fig-NN}
the value of $E_\pi^*$ increases monotonously with $p_0$, and
approaches its upper limit
$(m_N^2+m_{\pi}^2)/(2m_N)\cong 0.48$~GeV at $p_0\rightarrow \infty$.
Note that for $p_0=6$~GeV/c and $8.4$~GeV/c
used in p+A experiments \cite{baldin-74}  maximal values of the pion energy
in p+N reactions are equal to
$E_\pi^*\cong 0.38$~GeV and $0.41$~GeV, respectively.
These values are about two times smaller than those observed in the p+A
data \cite{baldin-74}. From Fig.~\ref{fig-NN} it also follows that at large
$p_0$, the value of $E_\pi^*$ becomes insensitive to $p_0$. Therefore, the production
of additional pions with not too large {\it positive} longitudinal momenta has small influence
on the $E_\pi^*$ value.

\begin{figure}[h!]
\centering
    \includegraphics[width=0.49\textwidth]{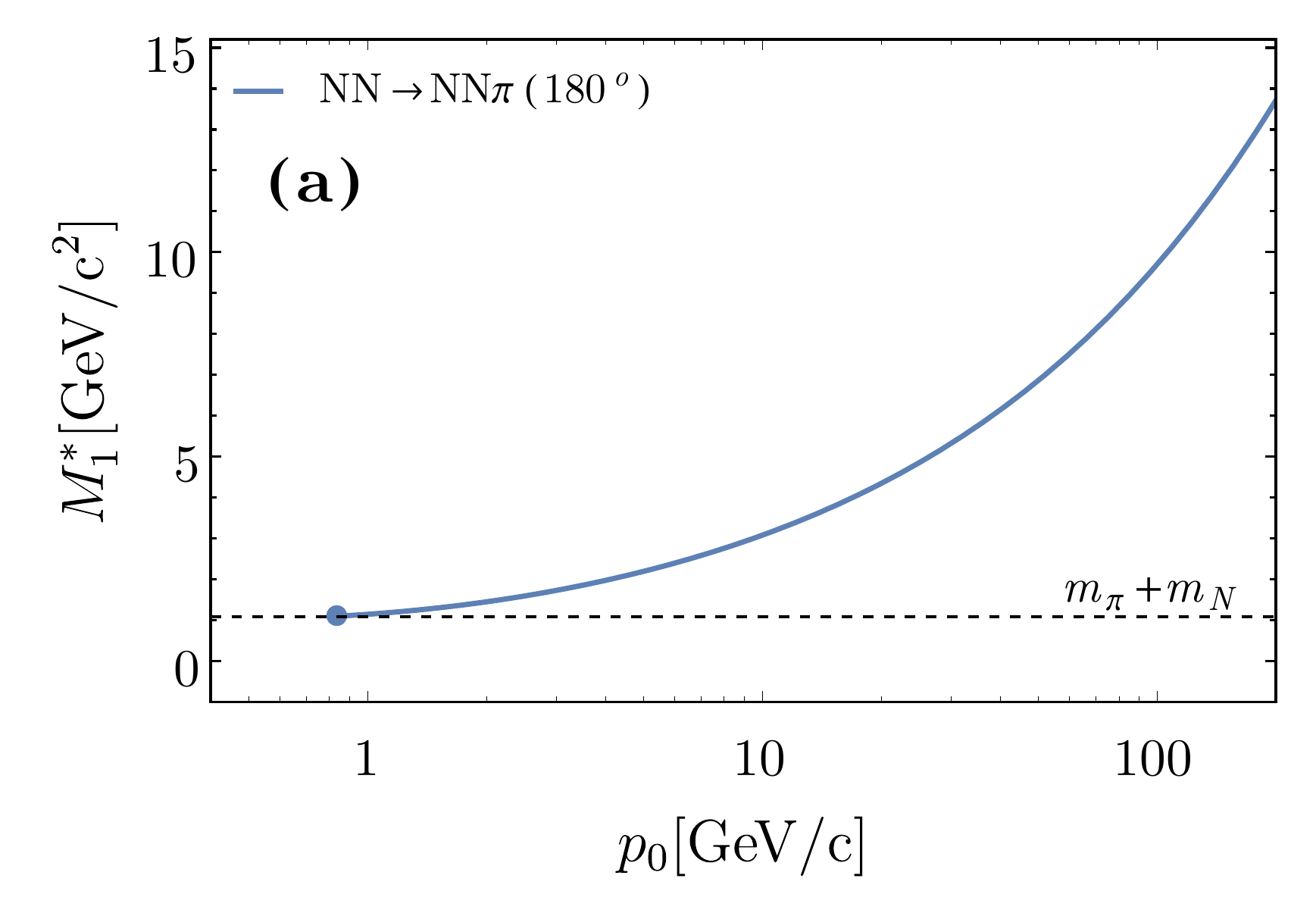}
     \includegraphics[width=0.49\textwidth]{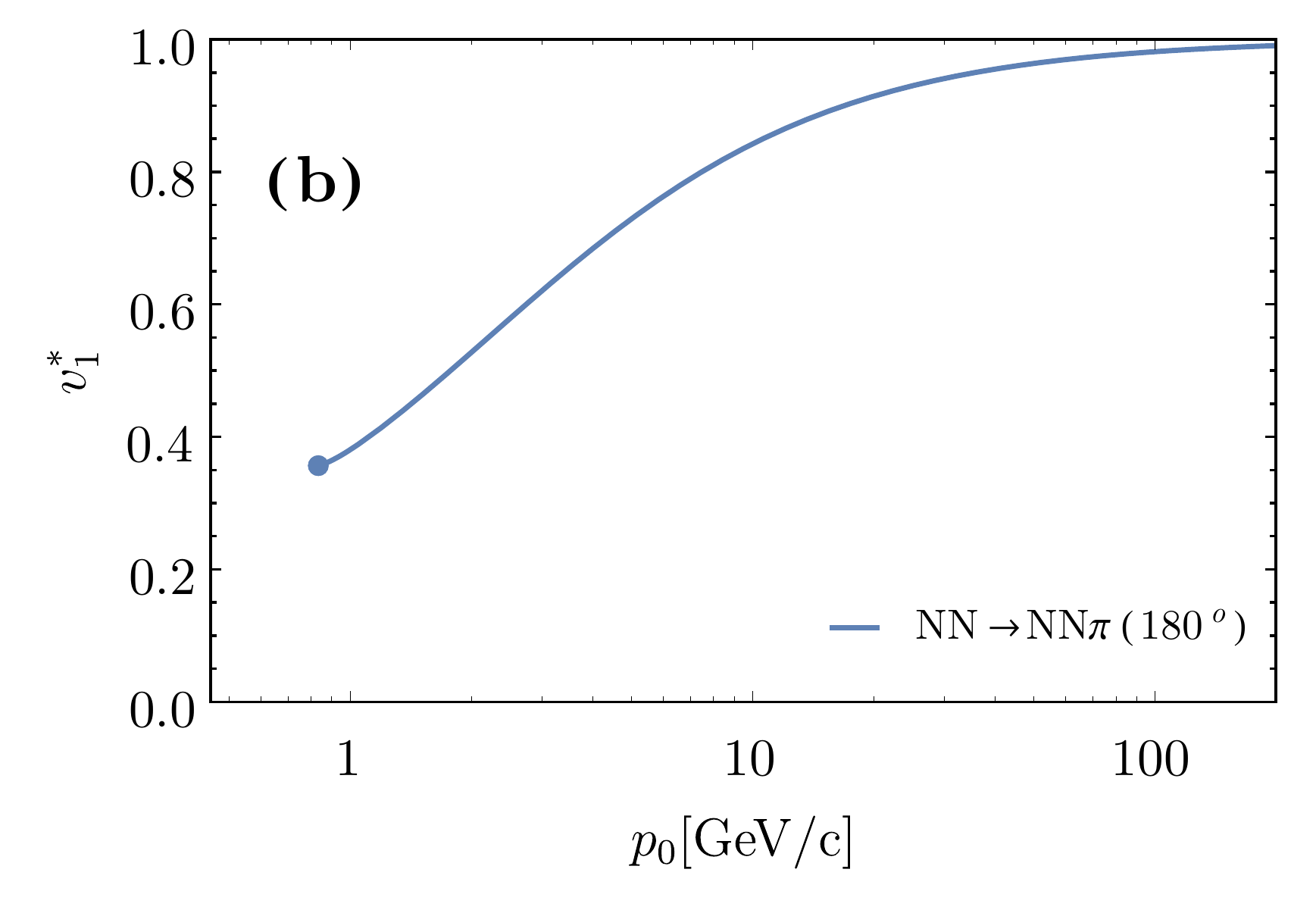}
    \caption{Invariant mass $M_1^*$ (a)  and longitudinal 
    velocity $v_1^*$ (b) of resonance $R$ from Fig.~\ref{fig-pN} (b)
    that decays into $\pi(180^\circ)$ with maximal energy $E^*_\pi$.
    $M_1^*$ and $v_1^*$ are calculated from Eq.~(\ref{M1*}) as functions of $p_{0}$. }
    \label{fig-M}
\end{figure}

The invariant mass $M_1^*$ and velocity $v_1^*$ of the resonance  $R$ that decays 
into $\pi(180^\circ)$ with maximal energy $E^*_\pi$ can be calculated as
\eq{\label{M1*}
M_1^* =\left[ \left(\sqrt{m_N^2+(p^*_1)^2}+E_\pi^* \right)^2-\left(p^*_1-p^*_\pi \right)^2\right]^{1/2}, ~~~~
v_1^*=\left[1- \frac{(M_1^*)^2}{(M_1^*)^2+(p^*_1-p^*_\pi )^2}\right]^{1/2}~,
}
and they are presented as functions of $p_0$ in Figs.~\ref{fig-M} (a) and (b), respectively.
We use the subscript 1 to denote the $M^*$ and $v^*$ values resulting from {\it single} p+N collision.

From Fig.~\ref{fig-M} (a) it is seen that $M_1^*$ increases monotonously with $p_0$,
and $M_1^*\cong \sqrt{m_Np_0}$ at $p_0\rightarrow \infty$.
Particularly, at $p_0\ge 20$~GeV/c one finds $M_1^*\ge 4.3$~GeV/c$^2$.
Thus, an evident question arises:
what is the physical origin of these massive systems?
Or, in other words, what is the microscopic
mechanism of a production of pions  at $180^{\circ}$
with energy $E_\pi^*$ in p+N reactions at large $p_0$.
In most relativistic transport models the objects that are responsible
for the production of new hadrons can be either resonances or
strings. In what follows we use a common notation, resonance,
for these objects and will
return to the discussion of their physical meaning in Sec.~\ref{sec-UrQMD}.

In a baryonic resonance decay,  $R\rightarrow {\rm N}+\pi(180^\circ)$,
the value of $E_\pi$
depends on the resonance mass $M_R$ and its longitudinal velocity $v_R$.
In the resonance rest frame
the pion energy and momentum can be easily found
\eq{\label{E0-R}
E^{0}_\pi= ~\frac{M_R^2-m_N^2+m_\pi^2}{2M_R}~,~~~~~p_\pi^{0}=\sqrt{(E_\pi^{0})^2-m_\pi^2}~.
}
The energy $E_\pi$, in the laboratory frame, is obtained as
\eq{\label{Epi-lab}
E_\pi= ~\frac{E^{0}_\pi - v_R p_\pi^{0}}{\sqrt{1-v_R^2}}~.
}
Therefore,
both the increase of $M_R$
and decrease of $v_R$ provide an extension of the available
kinematic region of $E_\pi$ for pions emitted at $180^{\circ}$.
This is most clearly seen if one neglects the pion mass, i.e., for $E_\pi^{0}\cong p_\pi^{0}$
(this is valid at $M_R\gg m_N$).
One then obtains
\eq{\label{Epi-app}
E_\pi\cong ~E^{0}_\pi \,\sqrt{\frac{1 - v_R}{1+v_R}}~.
}
Thus, the suppression of $E_\pi$  compared to $E^{0}_\pi$ can be interpreted as Doppler
effect (``red shift'' effect).

\begin{figure}[h!]
\centering
    \includegraphics[width=0.49\textwidth]{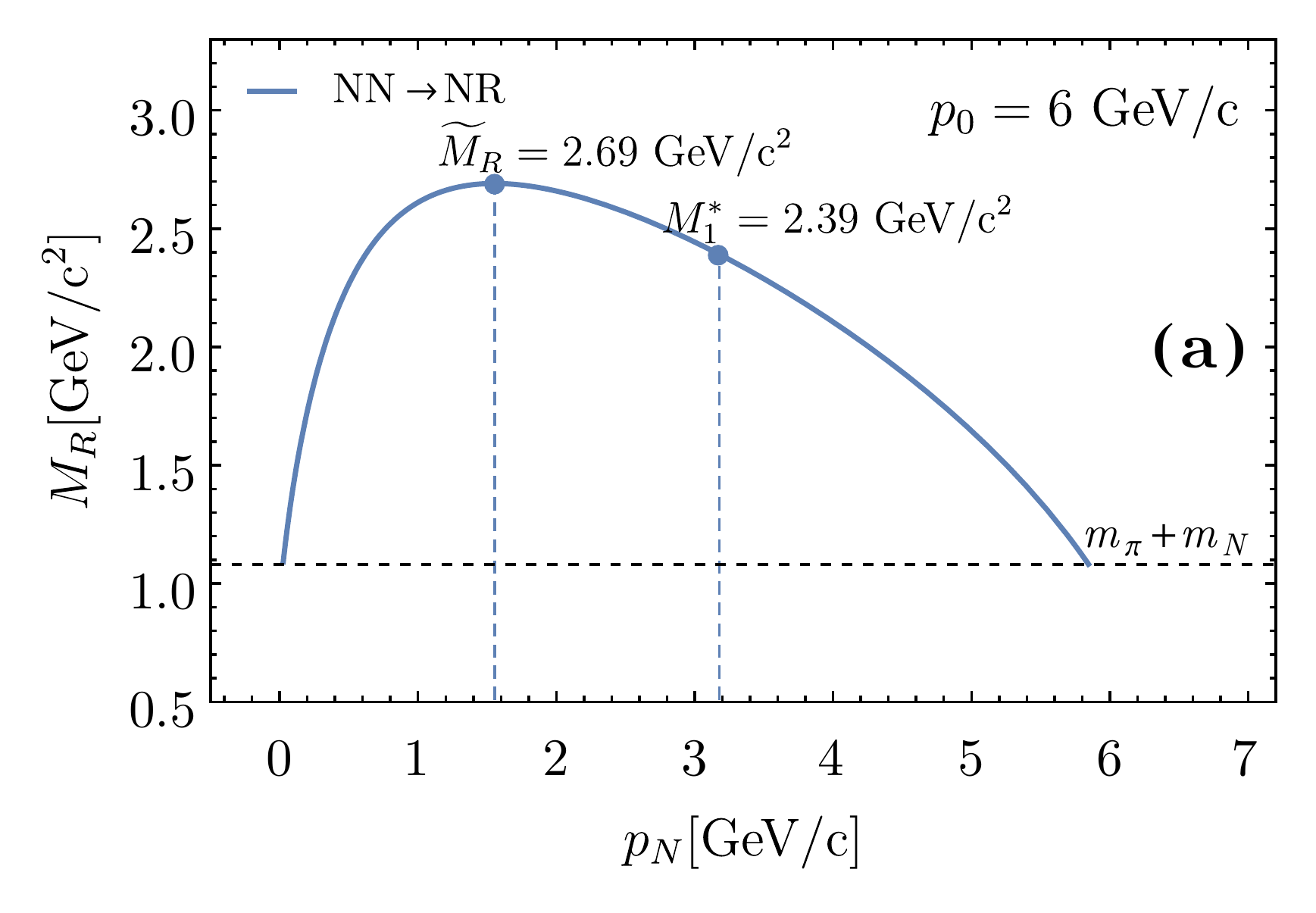}
     \includegraphics[width=0.49\textwidth]{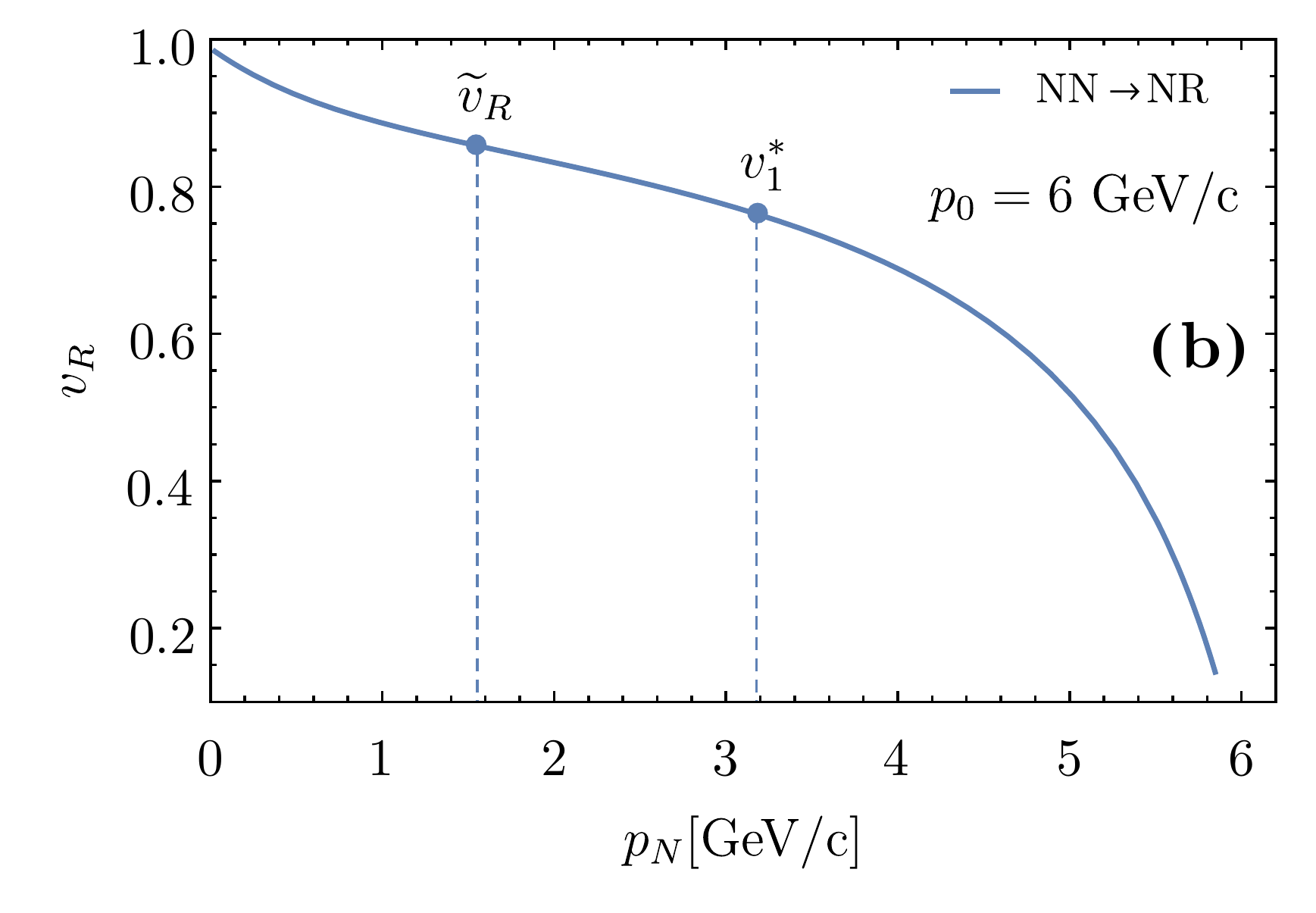}
    \caption{Resonance mass $M_R$ (a)
    and longitudinal velocity $v_R$ (b) that are possible 
    in NN$\rightarrow$NN$\pi(180^\circ)$ reactions presented in Fig.~\ref{fig-pN} (b).
    $M_R$ and $v_R$ are calculated from Eq.~(\ref{M-R}) and Eq.~(\ref{v-R}), respectively, as  functions of $p_N$ at fixed $p_0=6$~GeV/c.}
    \label{fig-Mp0}
\end{figure}

Let us consider the possible  values of $M_R$ and $v_R$ admitted by relativistic
kinematics in reaction
p+N$\rightarrow R$+N (see Fig.~\ref{fig-pN}) at fixed $p_0$.
As before, all particle momenta are
assumed to be directed along the collision axis.
From the energy-momentum conservation,
\eq{\label{conserv}
\sqrt{m_N^2+p_0}+m_N=\sqrt{M_R^2+p_R^2}+\sqrt{m_N^2+p_N^2}~,~~~~~~p_0=p_R+p_N~,
}
 one finds
\eq{\label{M-R}
M_R^2&=m_N^2+2p_{0}p_N-2\left(\sqrt{m_N^2+p_{0}^2}+m_N\right)\, \left(\sqrt{m_N^2+p_N^2}-m_N\right)~,\\
v_R&= \sqrt{1~-~\frac{M^2_R}{M_R^2+p_R^2}}~. \label{v-R}
}
The resonance mass $M_R$ (\ref{M-R}) and
velocity $v_R$ (\ref{v-R}) are  shown
in Figs.~\ref{fig-Mp0} (a) and (b), respectively,
as functions of $p_N$ at fixed $p_0=6$~GeV/c .

The maximum value of $M_R=\tilde{M}_R$ found from Eq.~(\ref{M-R}) at fixed $p_0$ corresponds
the condition $\partial M/\partial p_N=0$
that gives
\eq{\label{pN}
\tilde{p}_N\equiv\sqrt{\frac{m_N}{2}\left(\sqrt{m_N^2+p_{0}^2}-m_N\right)}~.
}
Note that Eq.~(\ref{pN}) is equivalent to the condition that the resonance with mass $M_R$
and nucleon move with the same velocities, both equal to $v_R$.
It should be  noted that the nucleon momenta $\tilde{p}_N$ given by Eq.~(\ref{pN})  is smaller
than the value of
$p^*_1=(p_0+p^*_\pi)/2$ given by Eq.~(\ref{p*1}) and required
to obtain $E_\pi = E_\pi^*$. One then finds that
$\tilde{M}_R$ obtained from Eqs.~(\ref{M-R}) and (\ref{pN}) is
larger than the invariant mass $M_1^*$ given by Eq.~(\ref{M1*})
that is needed to emit  $\pi(180^\circ)$ with maximal energy $E^*_\pi$.

\begin{figure}[h!]
\centering
    \includegraphics[width=0.49\textwidth]{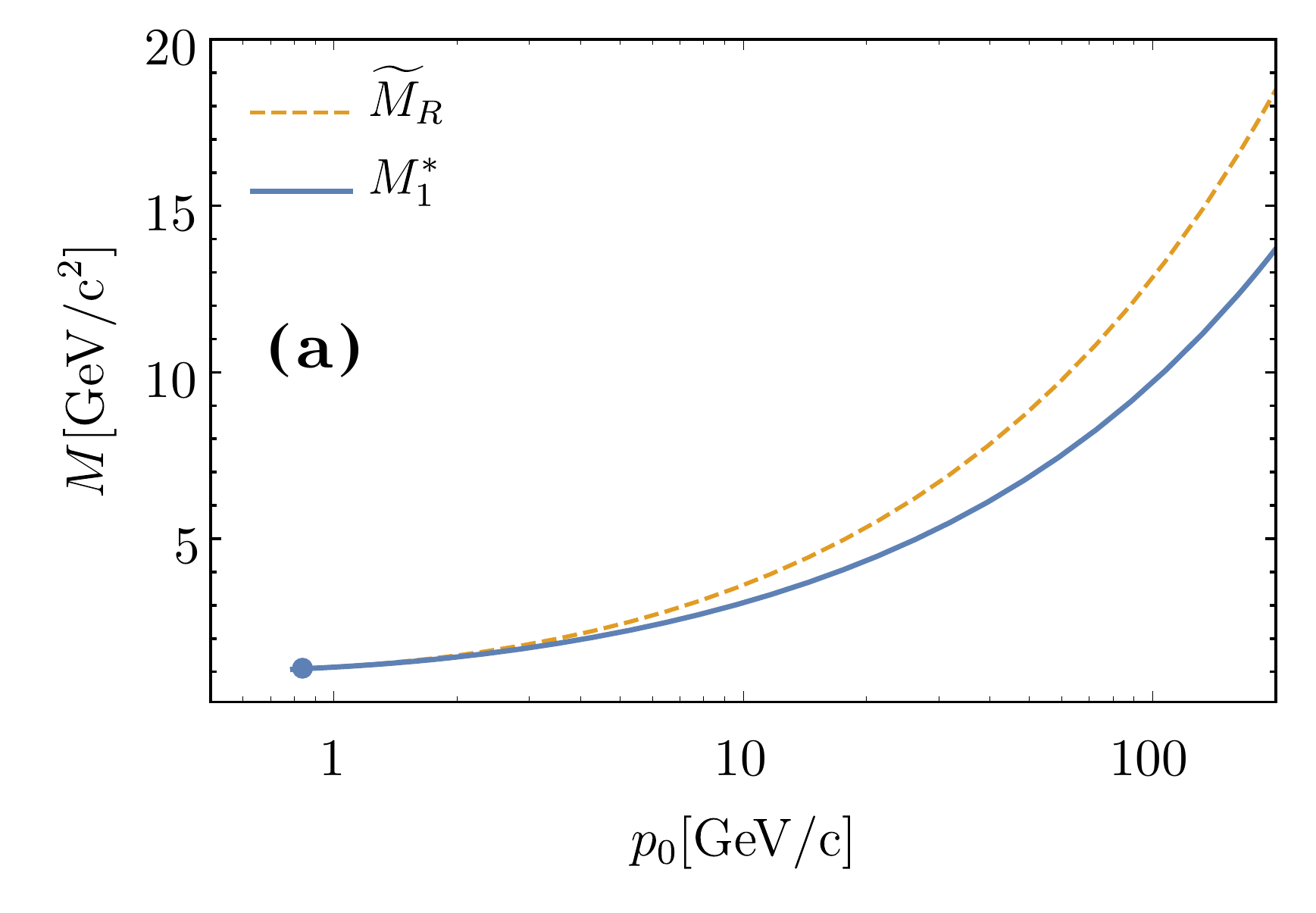}
    \includegraphics[width=0.49\textwidth]{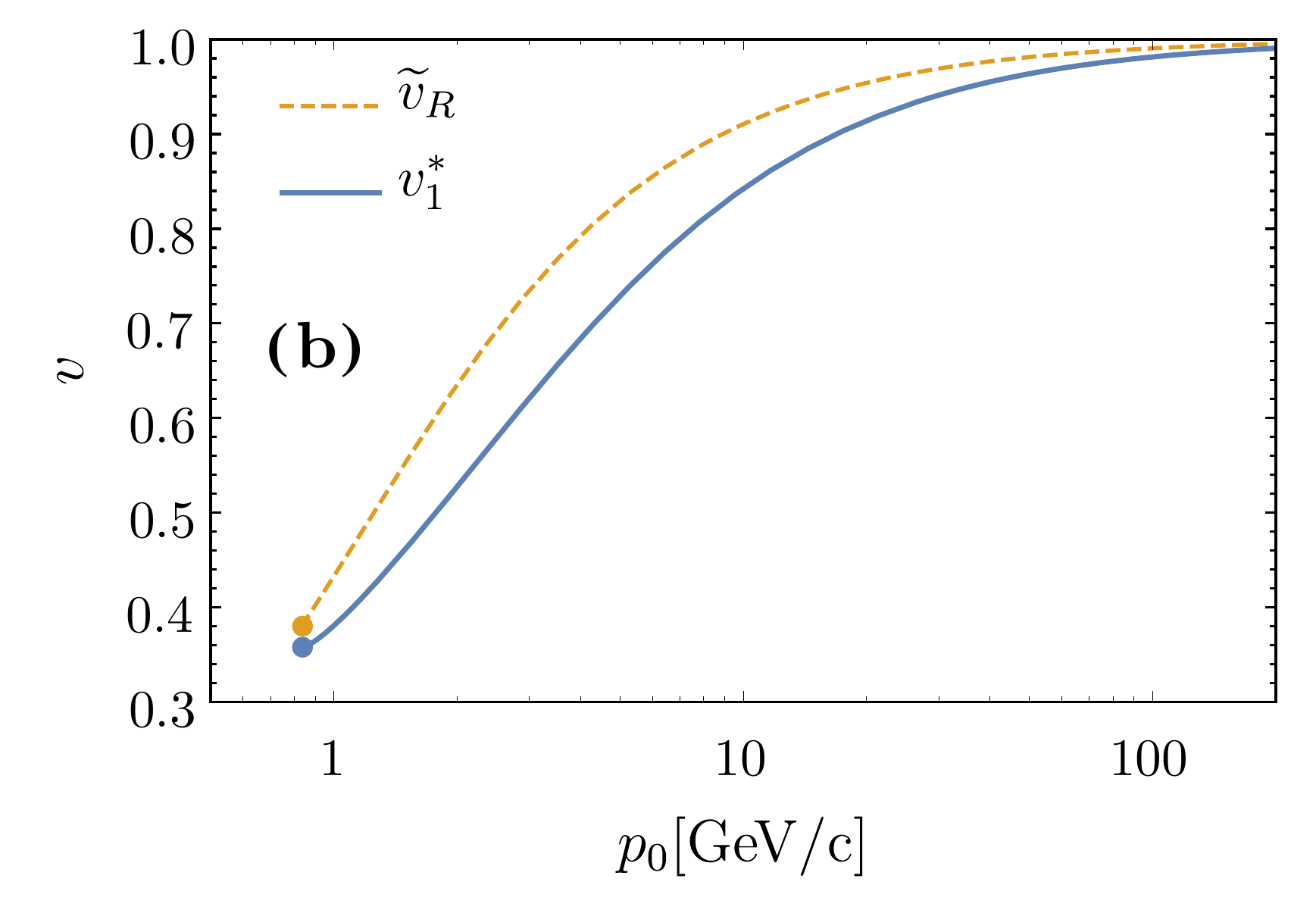}
    \caption{(a): Invariant mass $M_1^*$ given by Eq.~(\ref{M1*})
    and $\tilde{M}_R$ given by Eqs.~(\ref{M-R})
    and (\ref{pN})   are presented
    as functions of $p_0$ by solid and dashed line, respectively.
    (b): Velocity $v_1^*$ (\ref{M1*}) and $\tilde{v}_R$ (\ref{v-R},\ref{pN}) are shown
    as functions of $p_0$ by solid and dashed line, respectively.
    Values $M_1^*$ and $v_1^*$ provide the maximal energy $E_\pi^*$ of $\pi(180^\circ)$.
    $\tilde{M}$ and $\tilde{v}_R$ provide the maximal mass of intermediate resonance $R$ in 
    the reaction shown in Fig.~\ref{fig-pN} (b). } \label{fig-MM}
\end{figure}

One might intuitively
expect that maximal energy of $\pi(180^\circ)$
should correspond to a maximal mass $\tilde{M}_R$ of an intermediate resonance.
This is because  larger value of mass $M_R$ leads to a larger value of the pion energy
in its {\it rest frame} as given by Eq.~(\ref{E0-R}). Moreover,
a naive expectation might be that a resonance with a larger value of $M_R$ should move with a
slower velocity $v_R$ that makes $E_\pi(180^\circ)$ yet larger.
These intuitive expectations
are, however, not correct.
Conservation of both the energy and momentum leads to the
non-trivial connection between $M_R$ and $v_R$ values. As seen from Fig.~\ref{fig-Mp0} (b), $v_R$
decreases monotonously with $p_N$. Thus, the largest mass $\tilde{M}_R$ does not correspond
to the smallest velocity $v_R$.
On the other hand, too small velocity $v_R$
would correspond to too small resonance mass $M_R$, and the
backward production of cumulative pions would become impossible.
To emit a $\pi$-meson at $180^\circ$ with $E_\pi = E_\pi^*$
a compromise between large $M_R$ and small $v_R$ should be found,
i.e., the value of $E_\pi$ (\ref{Epi-lab}) should be maximized.
These {\it optimal}
values of $M_R$ and $v_R$  are just $M_1^*$ and $v_1^*$, respectively.
A comparison of $M_1^*$ with $\tilde{M}_R$ and $v^*$ with $\tilde{v}_R$
as functions of $p_0$ is presented in Fig.~\ref{fig-MM} (a) and (b), respectively.

\section{Successive collisions  with nuclear nucleons}\label{sec-multi}
As shown in the previous section, an energy
of the pion emitted at $180^{\circ}$ in p+N reactions is restricted by the
$E_\pi^*$ value presented in Fig.~\ref{fig-NN} as a function of $p_0$.
However, pion energies even larger than
$2E_\pi^*$ were observed experimentally
in p+A collisions \cite{baldin-74,baldin-77}.
One therefore should admit that the production of cumulative pions in p+A collisions
involves more than one target nucleon.

A multi-nucleon system inside a nucleus can manifest itself in various ways.
For example,  inter-nucleon interactions in a nucleus
may lead to one-particle momentum distribution with
a long tail (see Refs.~\cite{voloshyn,frankel,frankfurt,luk-79}). Another possibility
to create particles beyond the kinematical boundary allowed in p+N reactions
is to assume that a projectile proton interacts in an individual collision
simultaneously with a multi-nucleon system, the so-called ``grain'' or ``flucton''
(see Ref.~\cite{baldin-77,burov-77}). Indeed, if the target mass equals
$2m_N$, $3m_N$, etc., cumulative particle production may take place.
Both these approaches refer to some uncommon
aspects of nuclear physics. An object responsible for the cumulative particle production
is assumed  to exist inside a nucleus prior to its collision with a projectile.
Interpretation of these objects as multi-quark states
and calculations based on parton distribution functions
were discussed in
Refs.~\cite{burov-77,efremov,efremov-88,QCD}.

In the present study we will advocate the approach suggested in
~\cite{gor-77}, see also Refs.~\cite{gor,gor-80,gor-82,shmonin,kalinkin}.
It assumes that instead of a large mass of multi-nucleon target, cumulative
particle production takes place due to the large mass of the projectile
baryonic resonance created in the {\it first} p+N collision
and  propagated further through the nucleus.
This baryonic resonance  has a chance to interact with other nuclear nucleons
earlier than it decays to free final hadrons. As we will demonstrate
in this section the {\it successive collisions}
of the baryonic resonance with  nuclear nucleons
may both  enlarge $M_R$ and, simultaneously, reduce $v_R$ values in comparison to their
$M_1^*$ and $v_1^*$ values given by Eq.~(\ref{M1*}).
As seen from Eq.~(\ref{Epi-lab}), both effects
lead to larger values of $E_\pi$ and, thus,
extend
the kinematic region for cumulative pion production.
In contrast to the ``grain'' model,
the object responsible for the cumulative production of $\pi(180^\circ)$, i.e.,
the heavy and slow moving resonance, does not exist
inside a nucleus but is formed during the whole evolution process of p+A reaction.
Note that the role of rescattering effects in cumulative hadron production
were also discussed in Refs. \cite{re-1,re-2,re-3}.

\begin{figure}[h!]
\centering
    \includegraphics[width=0.65\textwidth]{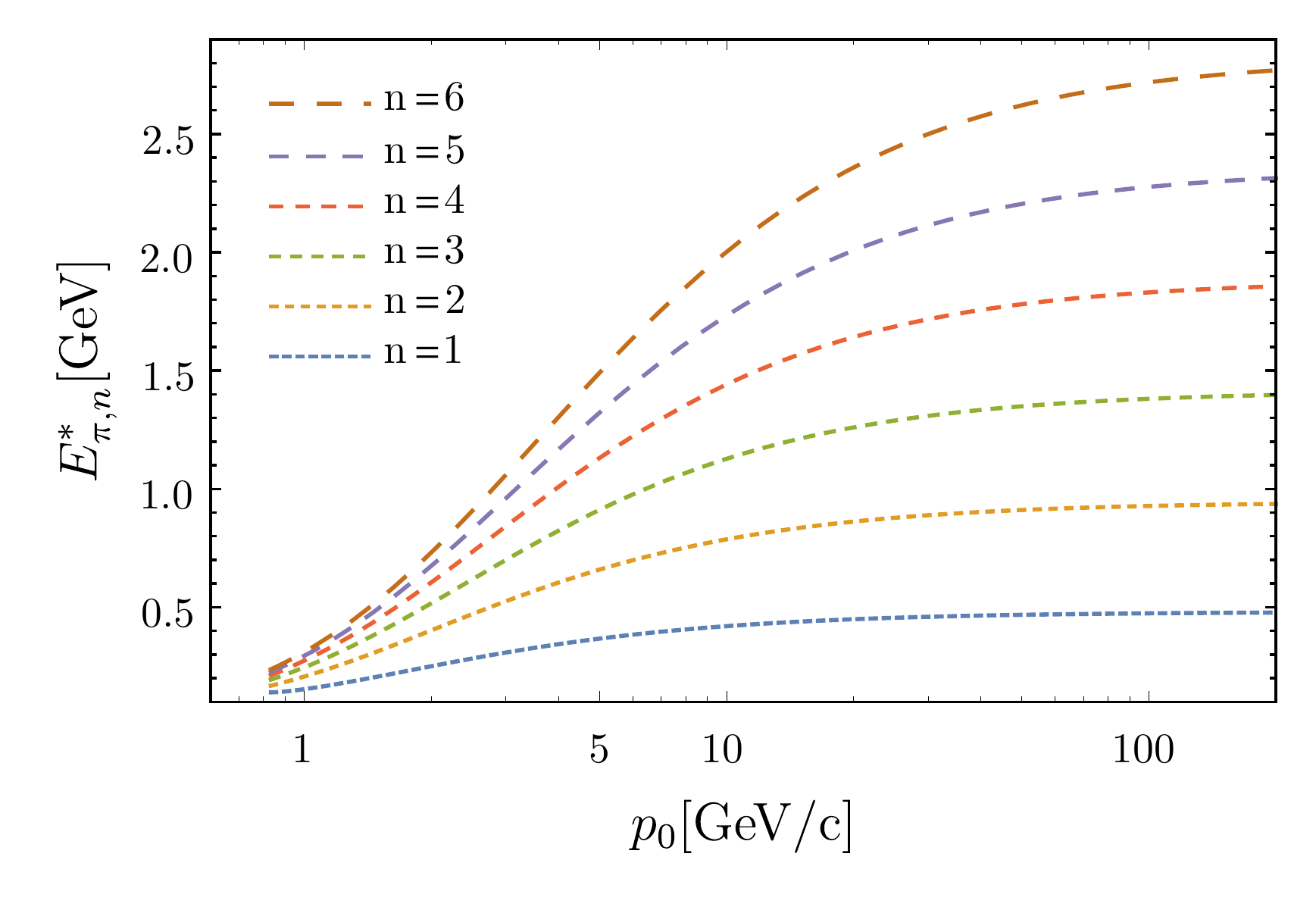}
    \caption{Maximal energies $E_{\pi,n}^*$ (\ref{E-max-n}) of $\pi$-meson emitted at
    $180^{\circ}$ after $n$ successive collisions with nuclear nucleons ($n=1,\ldots,6$) as  functions of
    projectile proton momentum $p_0$.  }
    \label{fig-En}
\end{figure}

Let us consider successive collisions with nuclear nucleons:
${\rm p}+{\rm N}\rightarrow R_1+ {\rm N}$, $R_1+{\rm N}\rightarrow R_2+ {\rm N}$,
{\it etc.}
It is assumed that
after $n$-th collision the baryonic resonance decays, $R_n\rightarrow \pi(180^\circ)$ + N.
The energy and momentum conservation between initial and final state read as
\eq{\label{cons-n}
\sqrt{m_N^2+p_0^2}+n\,m_N =\sum_{i=1}^{n+1}\sqrt{m_N^2+p_i^2}+E_\pi~,~~~~~~
p_{0}=\sum_{i=1}^{n+1}p_i-p_\pi~.
}
The maximal pion energy $E_\pi$ after $n$ successive collisions 
denoted now $E_{\pi,n}^*$ can be found
from  Eq.~(\ref{cons-n}) using the extremum conditions
$\partial E_\pi /\partial p_{i}=0$. This leads to
\eq{\label{p*n}
p_{N,n}^*\equiv p_1=p_2=...=p_{n+1}=\frac{p_0+p^*_{\pi,n}}{n+1}~,
}
and gives an implicit equation for $E_{\pi,n}^*$
\eq{\label{E-max-n}
E_{\pi,n}^*=n\,m_N+\sqrt{m_N^2+p_0^2}-
%\left(n+1\right)\,\sqrt{m_N^2+(p^*_{N,n})^2}}
\left(n+1\right)\,\sqrt{m_N^2+\left(\frac{p_0+
\sqrt{(E_{\pi,n}^*)^2-m_{\pi}^2}}{n+1}\right)^2}~.
}
The maximal energies $E_{\pi,n}^*$ of pions emitted at $180^{\circ}$ are
presented in Fig.~\ref{fig-En}.

\begin{figure}[h!]
\centering
    \includegraphics[width=0.49\textwidth]{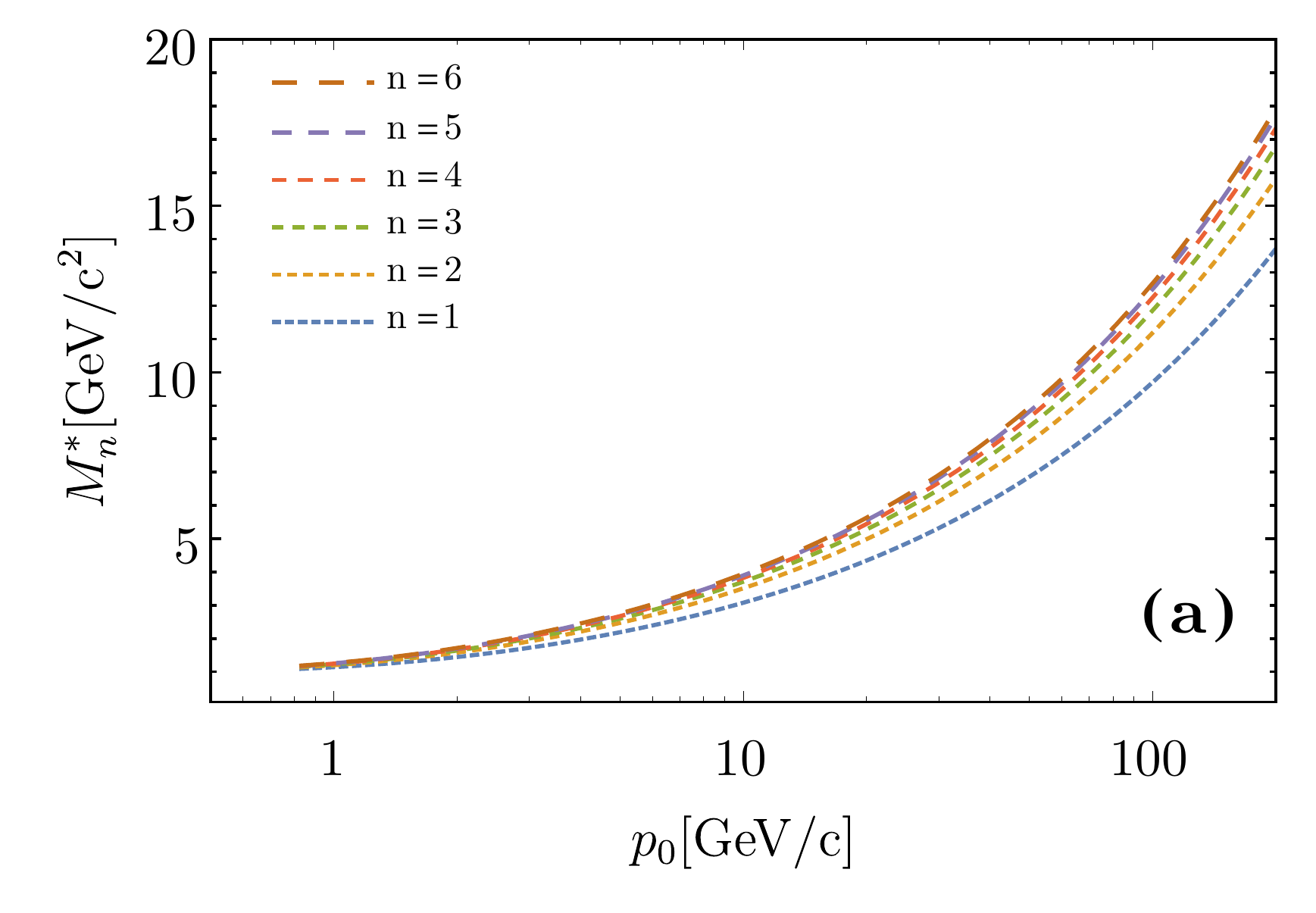}
       \includegraphics[width=0.49\textwidth]{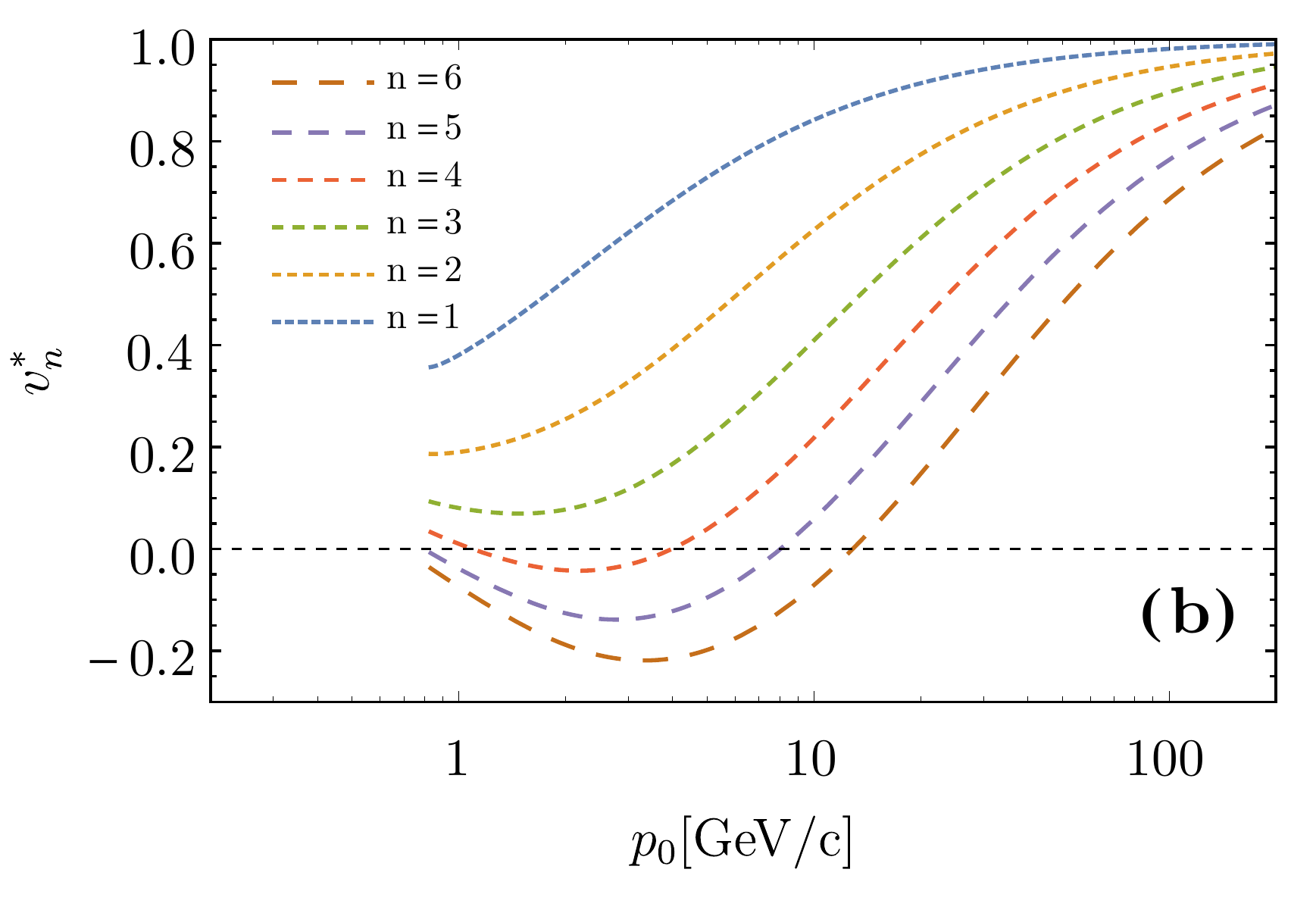}
    \caption{Invariant masses $M_n^*$ (a) and velocities $v_n^*$ (b)
    of the baryonic resonances after $n$ successive collisions with nuclear nucleons. 
    The values of $M_n^*$ and $v_n^*$ are required to provide the maximal energy
    $E_{\pi ,n}^*$ of $\pi(180^\circ)$.
    They are calculated from (\ref{Mn*}) and (\ref{vn*}), respectively,  as  functions of
    projectile proton momentum $p_0$. }
    \label{fig-Mn}
\end{figure}
Note that  for $n=1$ Eqs.~(\ref{p*n}) and (\ref{E-max-n}) are reduced
to Eqs.~(\ref{p*1}) and (\ref{E-max}), respectively, i.e.,
$E_{\pi,1}^*\equiv E^*_\pi$ which is also shown in Fig.~\ref{fig-NN}.
As seen from Fig.~\ref{fig-En},
the value of maximum pion energy $E_{\pi,n}^{*}$ in the backward
direction  increases essentially with the number of collisions $n$.
Besides, at any $n$, the value of $E_{\pi,n}^{*}$ increases monotonously with $p_0$, and
at $p_0\rightarrow \infty$
it goes to the upper limits
$n(m_N^2+m_{\pi}^2/n^2)/(2m_N)\cong n\cdot 0.48$~GeV, i.e., $E_{\pi,n}^*$
increases approximately linearly
with the number of collisions at $p_0\rightarrow \infty$.

The invariant mass $M^*_n$ and longitudinal velocity $v_n^*$ of the resonance
after $n$ successive collisions with nuclear nucleons are
given by the same formulae as in Eq.~(\ref{M1*}) but with a substitution of $p_1^*$ (\ref{p*1})
by $p_n^*$ (\ref{p*n})
\eq{\label{Mn*}
M_n^* &=\left[ \left(\sqrt{m_N^2+(p^*_n)^2}+E_{\pi,n}^* \right)^2-\left(p^*_{N,n}-p^*_{\pi,n} \right)^2\right]^{1/2}~, \\
|v_n^*|&=\left[1- \frac{(M_n^*)^2}{(M_n^*)^2+(p^*_{N,n}-p^*_{\pi,n} )^2}\right]^{1/2}~.\label{vn*}
}
These quantities as functions of $p_0$
are presented in Fig.~\ref{fig-Mn} for $n=1,\ldots,6$.
%Note that $M_1^*=M^*$ and $v_1^*=v^*$, which
%are presented in Fig.~\ref{fig-Mp0}.

Note also that Eqs.~(\ref{p*n}) and (\ref{E-max-n})  can be interpreted as a
collision of the projectile proton with the $n$-nucleon ``grain'', i.e.,
the resulting conservation laws for the cumulative production of $\pi(180^\circ)$ with
$E_\pi=E^*_{\pi,n}$
due to successive collisions with $n$ nucleons
look the same as in a collision with the $n$-nucleon ``grain''.

From Fig.~\ref{fig-Mn}  one sees that $M_n^*$ increases with $p_{0}$
and $M_n^*\cong \left(\dfrac{2n}{n+1}m_Np_0\right)^{1/2}$ at $p_0\rightarrow \infty$. Besides, the mass $M_n^*$
increases monotonously and velocity $v_n^*$ decreases monotonously with
the number of collisions $n$ at each $p_0$.
Note that $M_n^*$
%in contrast with pion energy $E_{\pi,n}^*$,
demonstrates only a modest increase with $n$.
Thus, the increase of $E_{\pi,n}^*$ with $n$
seen in Fig.~\ref{fig-En} takes place mainly due to a noticeable decrease of $v_n^*$ as it is seen from
Fig.~\ref{fig-Mn} (b).

A surprising behavior with $v_n^*<0$ for $n\ge 4$ is observed
at some finite regions of projectile momentum $p_0$, i.e., heavy resonance may start
to move backward after a large number of successive collisions for not too large $p_0$.
In p+N$\rightarrow R$+N reactions the only $v_R$ values with $v_R>0$ are permitted.
This was tacitely assumed in Eqs.~(\ref{Epi-lab}) and (\ref{Epi-app}). However,
%if $v_R<0$ appears after several successive collisions,
Eqs.~(\ref{Epi-lab}) and (\ref{Epi-app})
remain also valid for $v_R< 0$. In contrast to the case when $E_\pi$ is reduced
due to $v_R>0$,
the $v_R<0$ values lead to enhancement of $E_\pi$.
Particularly, Eq.~(\ref{Epi-app})
exhibits the ``blue shift'' of the $\pi(180^\circ)$ energy at $v_R<0$.

\begin{figure}[h!]
\centering
    \includegraphics[width=0.49\textwidth]{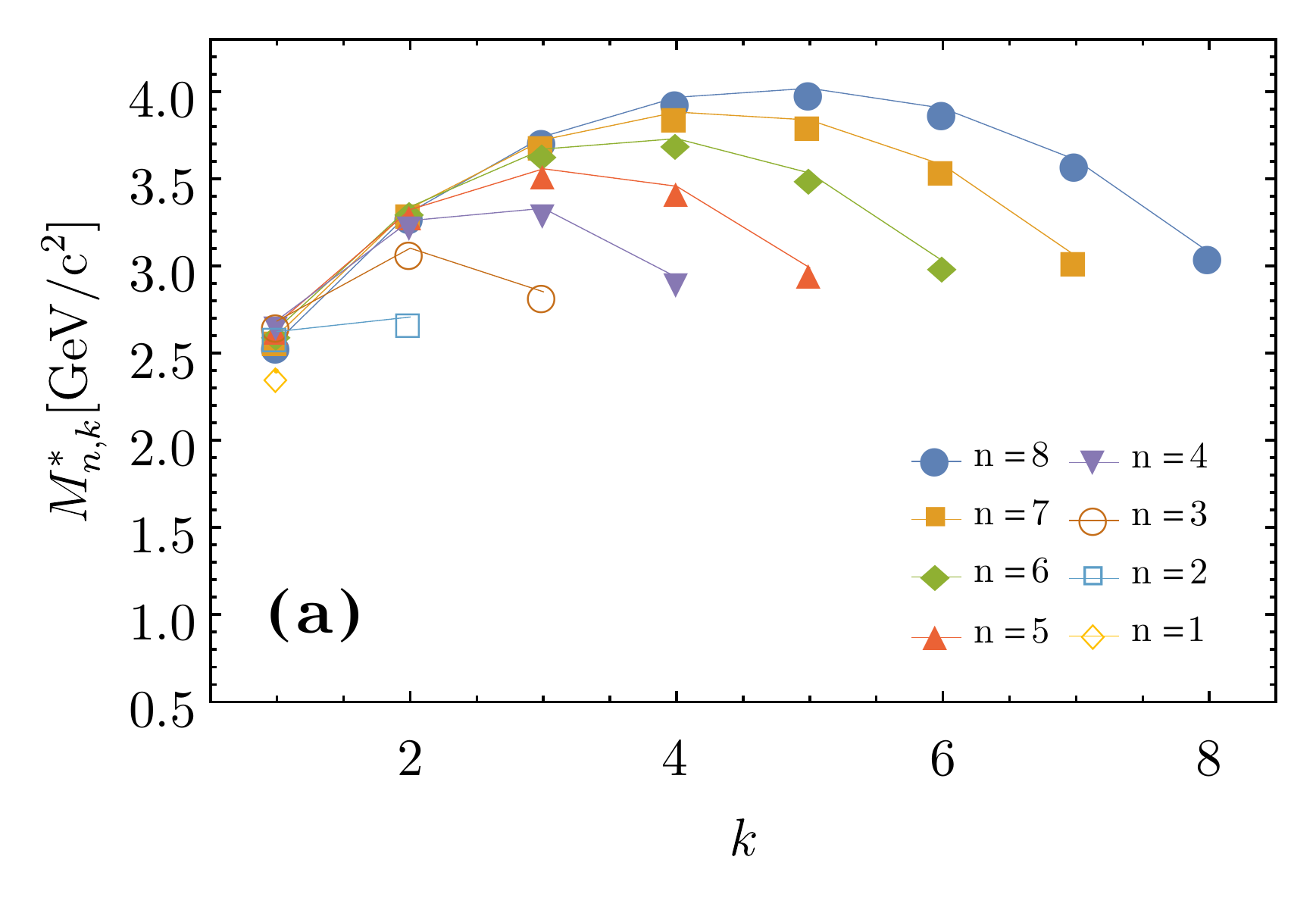}
    \includegraphics[width=0.49\textwidth]{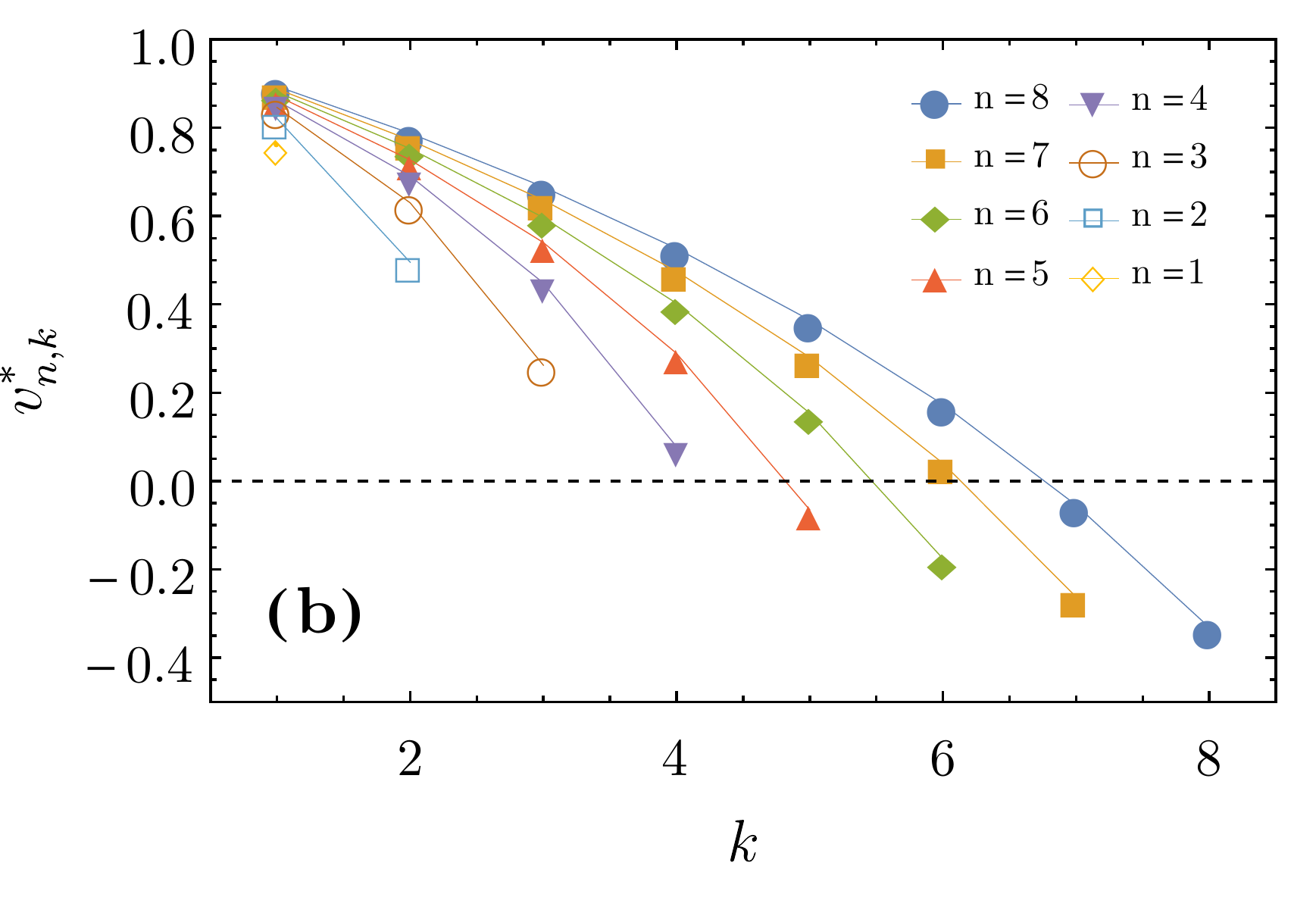}
    \caption{Mass of resonance $M^*_{n,k}$ (a) and its velocity $v^*_{n,k}$ (b)
     for $k$-th collision $k=1,\ldots,n$ of total $n=1,\ldots,8$ at fixed $p_0=6$~GeV/c.}
    \label{fig-Mkn}
\end{figure}

Let us look at the evolution of the resonance mass and its velocity
due to successive collisions with nuclear nucleons, namely,
how the values of $M_R$ and $v_R$ should evolve
after each $k$th collisions ($k=1,\ldots,n$) to reach their final values
$M_n^*$ and $v_n^*$ after the $n$th collision.  We calculate the masses $M^*_{n,k}$
and velocities $v^*_{n,k}$ of the resonance after each of $n$ successive collisions, $k=1,\ldots,n$.
The corresponding masses  $M^*_{n,k}$
and velocities $v^*_{n,k}$ are presented in Fig.~\ref{fig-Mkn} (a) and (b),
respectively, as functions of $k=1,\ldots,n$ for different $n=1,\ldots,8$,
and at fixed $p_0=6$~GeV/c. Note that $M_{n,n}\equiv M_n^*$ and $v_{n,n}\equiv v_n^*$,
and these quantities are shown in Figs.~\ref{fig-Mn} (a) and (b), respectively.

One sees from Fig.~\ref{fig-Mkn} (a) that at $k\approx n/2$ the mass $M^*_{n,k}$
reaches its maximum and starts to decrease at $k\ge n/2$ in each subsequent collision.
On the other hand, the resonance velocity decreases monotonously after each subsequent collision
and may become negative at large $k$.
%{\bf It seems that only $v^*_{n,n}$ may become negative!!}

\begin{figure}[h!]
\centering
    \includegraphics[width=1\textwidth]{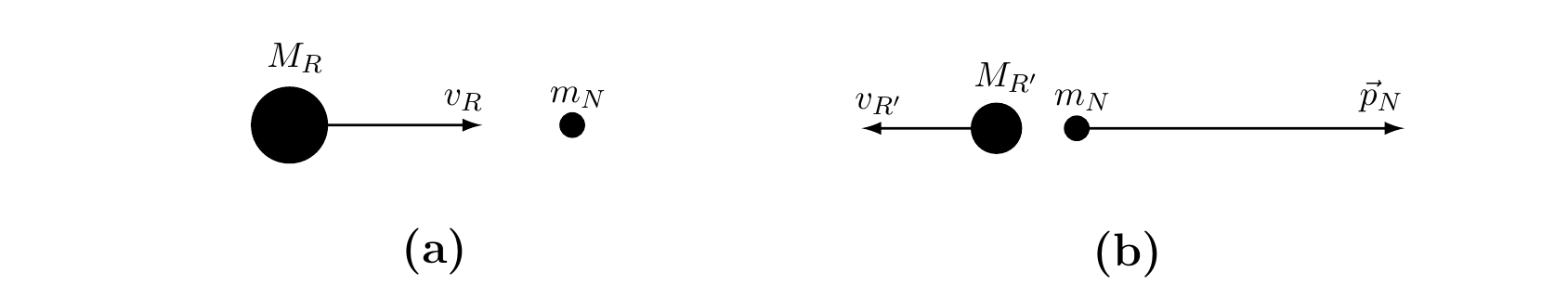}
    \caption{$R$+N$\rightarrow R'$+N reaction. (a): initial stage.  (b): final stage.
      }
    \label{fig-v}
\end{figure}

We demonstrate now explicitly that a massive particle (resonance) may change its  motion
in backward direction after a collision with a lighter particle (nucleon) at rest.
This is schematically shown in Fig.~\ref{fig-v}.

\begin{figure}[h!]
\centering
    \includegraphics[width=0.49\textwidth]{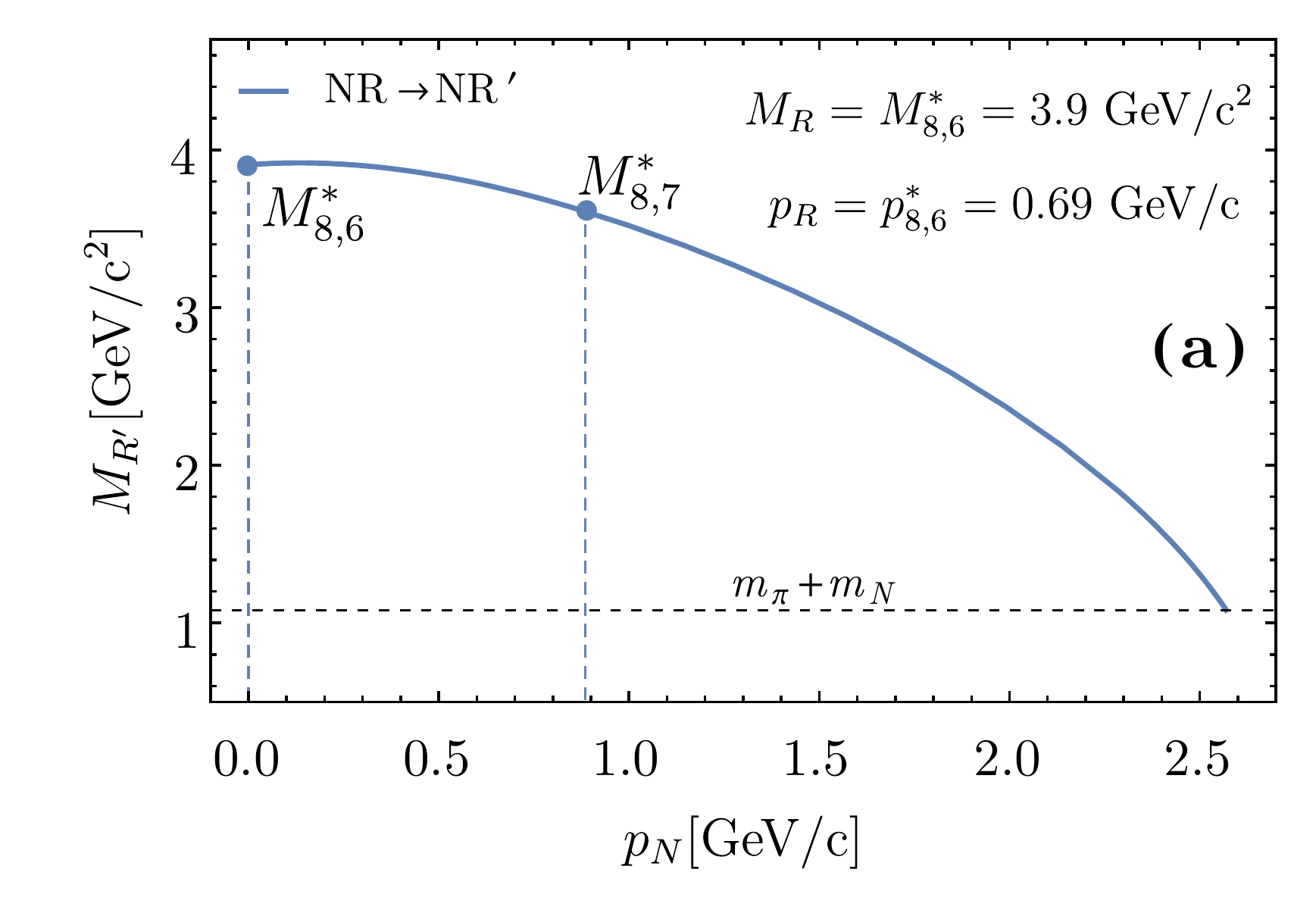}
     \includegraphics[width=0.49\textwidth]{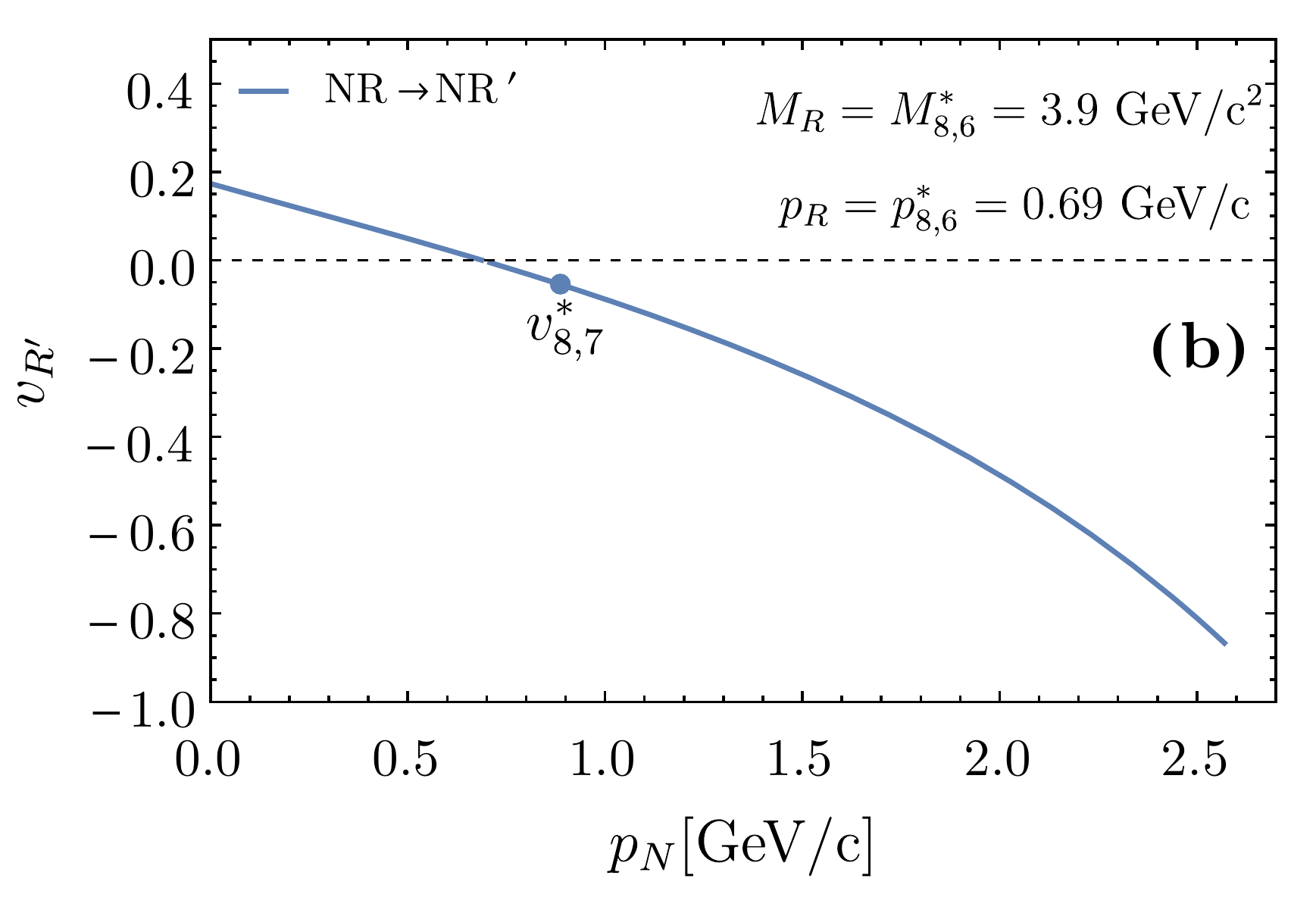}
    \caption{The resonance mass $M_{R'}$ (a) given by Eq.~(\ref{RR})
    and its velocity $v_{R'}$ (b) for reaction on Fig.~\ref{fig-v} calculated  at  $M_R=M^*_{8,6}=3.9$~GeV$\mathrm{/c^2}$ and $v_R=v_{8,6}^*=0.17~$.}
    \label{fig-RR}
\end{figure}
This is only possible due to the relativistic kinematics which admits the changes of particle masses
in a course of the reaction.
Let us consider the reaction
$R+{\rm N}\rightarrow R'+{\rm N}$. The conservation laws for this reaction look similar
to Eq.~(\ref{conserv})
\eq{\label{RR}
\sqrt{M_R^2+p^2_R}+m_N =\sqrt{M_{R'}^2+p_{R'}^2}+\sqrt{m_N^2+p_N^2}~,~~~~~~p_R=p_{R'}+p_N~,
}
but the projectile mass is now $M_R>m_N$.
To be definite, we take the values of $M_{R}=M^*_{8,6}=3.9$ GeV$\mathrm{/c^2}$  and $v_{R}=v_{8,6}^*=0.17$
shown in Fig.~\ref{fig-Mkn}.
The possible values of $M_{R'}$ and $v_{R'}$ as functions of $p_N$
are shown in Figs.~\ref{fig-RR} (a) and (b), respectively.

It can be easily seen from Fig.~\ref{fig-RR} that negative values $v_{R'} <0$
become indeed  possible, but only if the resonance loses its mass, i.e., $M_{R'}< M_R$,
in a collision with a nucleon.

\begin{figure}[h!]
\centering
    \includegraphics[width=1\textwidth]{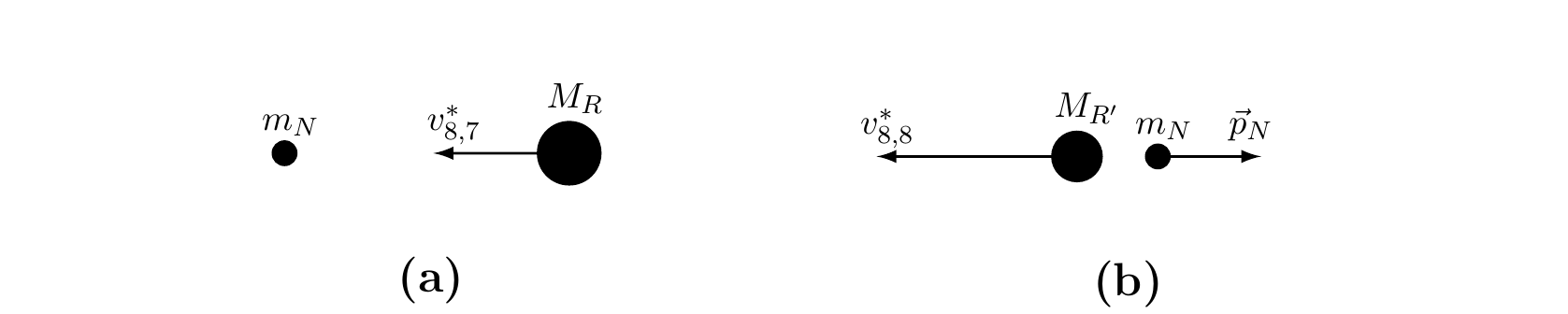}
    \caption{$R$+N$\rightarrow R'$+N reaction. (a): initial stage.  (b): final stage.
      }
    \label{fig-vv}
\end{figure}

\begin{figure}[h!]
\centering
    \includegraphics[width=0.49\textwidth]{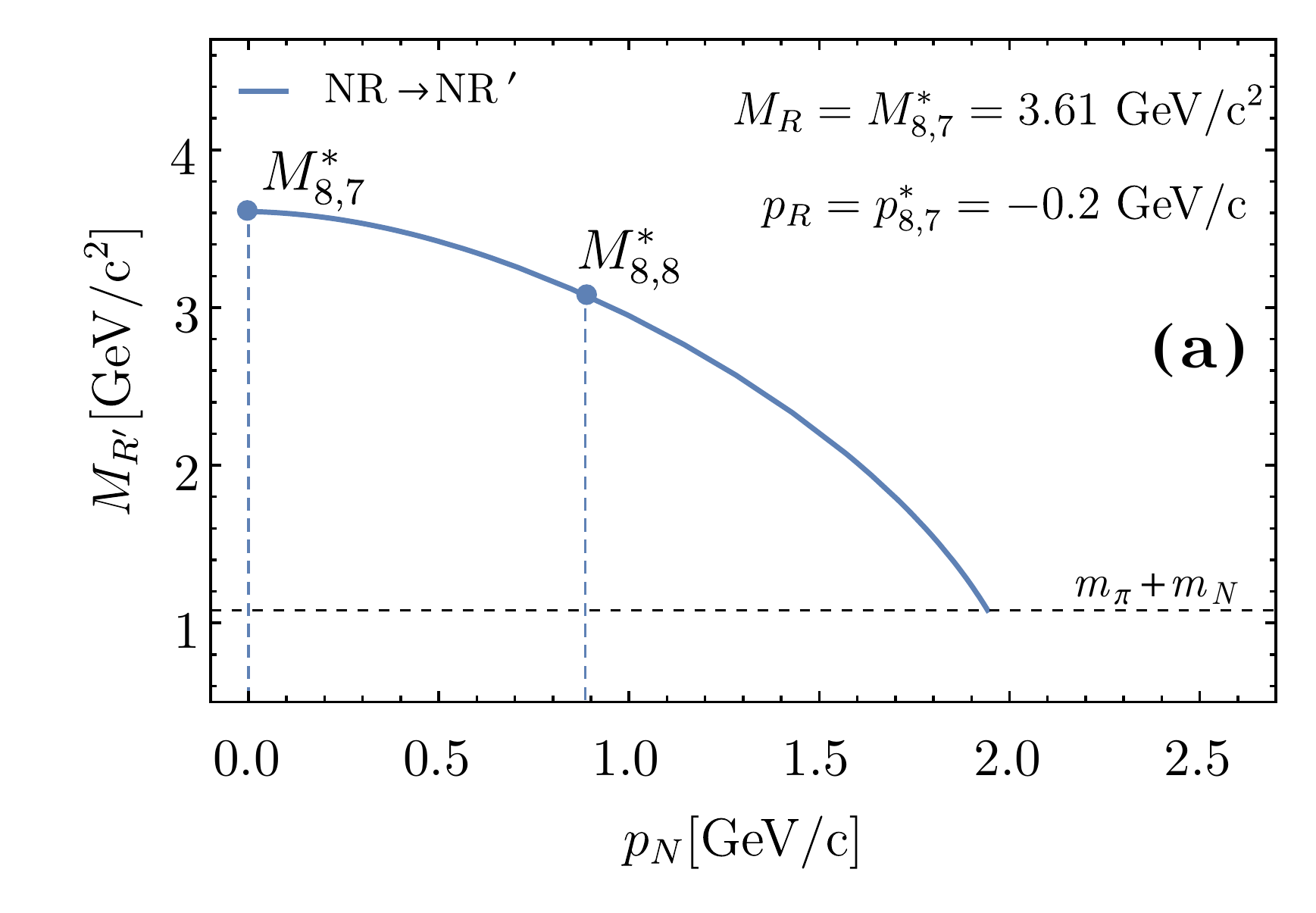}
     \includegraphics[width=0.49\textwidth]{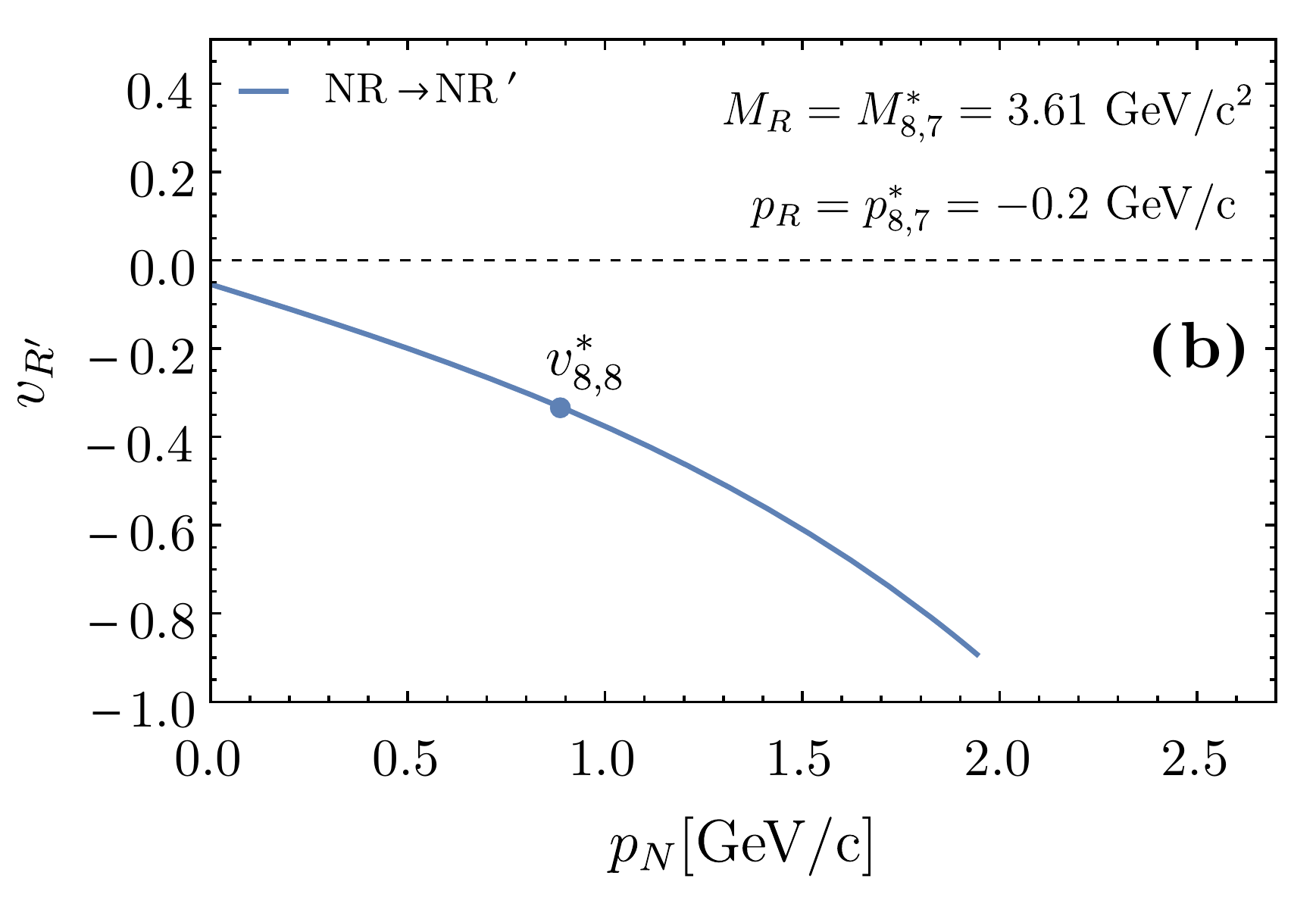}
    \caption{The resonance mass $M_{R'}$ (a) given by Eq.~(\ref{RR})
    and its velocity $v_{R'}$ (b) for reaction on Fig.~\ref{fig-vv} calculated  at  $M_R=M^*_{8,7}=3.61$~GeV$\mathrm{/c^2}$ and $v_R=v_{8,7}^*=-0.05$.
    }
    \label{fig-RRv}
\end{figure}

Taking initial values in Eq.~(\ref{RR}) as $M_{R}=M^*_{8,7}=3.61$ GeV$\mathrm{/c^2}$  and $v_{R}=v_{8,7}^*=-0.05$
one finds another unexpected possibility shown schematically in Fig.~\ref{fig-vv}.
In this case, the possible values of $M_{R'}$ and $v_{R'}$ as functions of $p_N$
% at fixed $p_R=\tilde{p}_R(p_0=6$~GeV/c$)=1.55$~GeV/c
are shown in Figs.~\ref{fig-RRv} (a) and (b), respectively.

It should be noted that the values of $M^*_n$ and $v^*_n$ found in this section
are by no means typical (or average) ones. In fact, the probability to reach these values
in p+A reaction is very small. In other words, cumulative pion production is a very rare process.

\section{UrQMD simulations}\label{sec-UrQMD}
In the present section we analyze the cumulative production of $\pi (180^\circ)$
within the UrQMD model~\cite{urqmd}. The UrQMD is a microscopic transport model
used to simulate relativistic heavy ion collisions in a wide range
of collision energies.
The version UrQMD-3.4 is used in the present study.
There are two distinct sources of particle production in the UrQMD:
formation and decays of resonances and formation and decays of hadronic strings.
Most important resonances for the cumulative pion production are the baryonic resonances
$N^*=(N^{0},N^+)$
%with $M=1440$~MeV,$\ldots$, 2250~MeV,
and $\Delta=(\Delta^-,\Delta^{0},\Delta^+,\Delta^{++})$.
%with $M=1232$~MeV,$\ldots$, 1950~MeV.
With increasing of  $p_0$ the excitations of baryonic strings
open the new channels of hadron production in p+N reactions. At
projectile momenta
$p_0 \ge 18$~GeV/c the string production dominates in the UrQMD
description of the inelastic p+N cross section \cite{urqmd}.

All type of binary collisions (\ref{reac}), particularly  $R+{\rm N}\rightarrow R'+{\rm N}$,
considered in the previous sections are possible
in the UrQMD model. The UrQMD has an upper limit for the mean resonance mass
(this limit is different in different versions of he model). 
However, due to continuous Breit-Wigner mass distribution quite big resonance masses may appear.
The masses of the strings are not restricted. On the other hand, in the UrQMD, as well as in
other relativistic transport model, the reactions ${\rm String}+h \rightarrow X $ are not permitted.
This means that, in contrast to resonances, strings can not participate in secondary reactions
as real objects, only their decay products can take place in successive reactions.

\subsection{UrQMD simulation of p+p reactions}
\begin{figure}[h!]
\centering
%\begin{floatrow}
\includegraphics[width=0.48\textwidth]{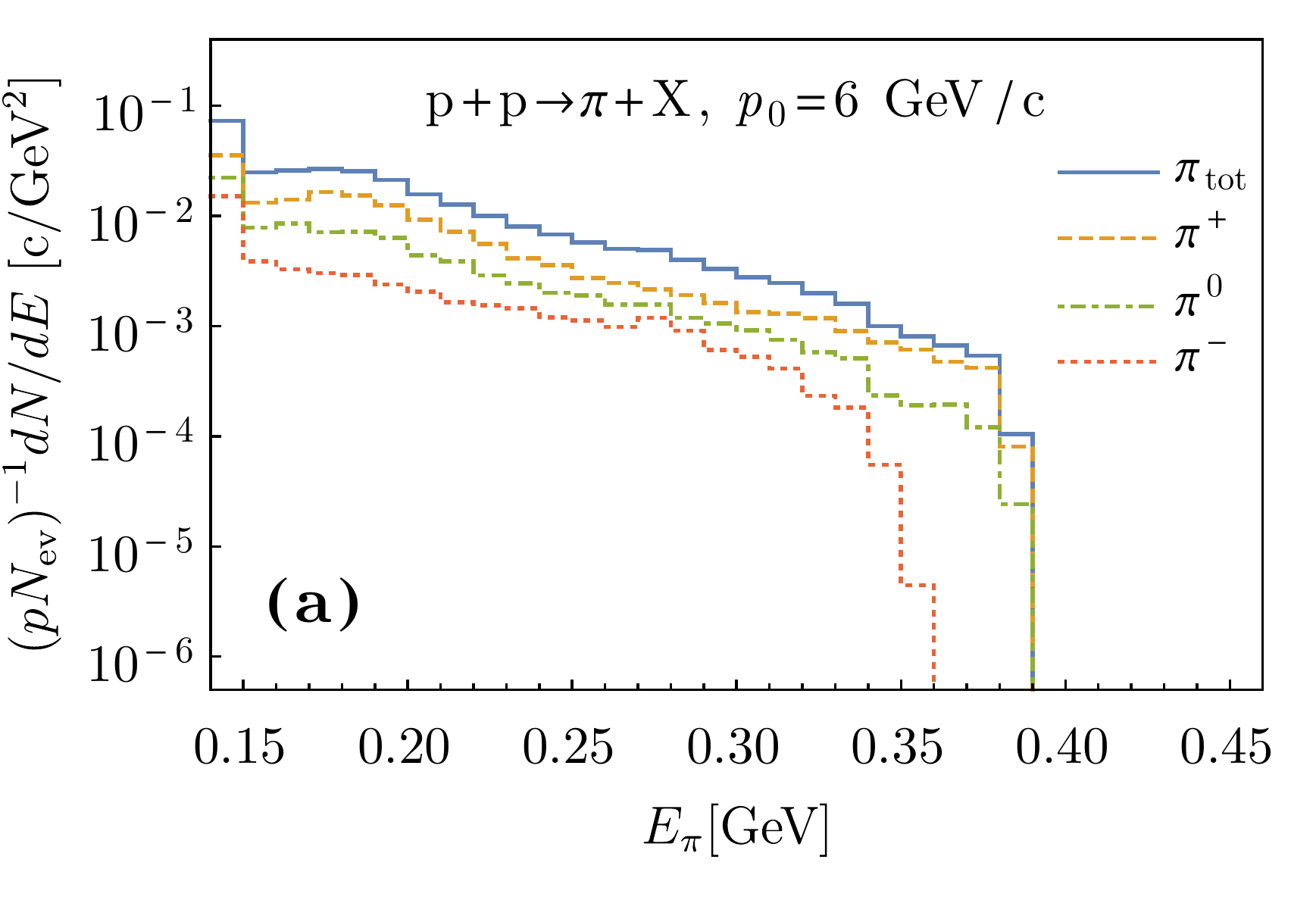}
\includegraphics[width=0.48\textwidth]{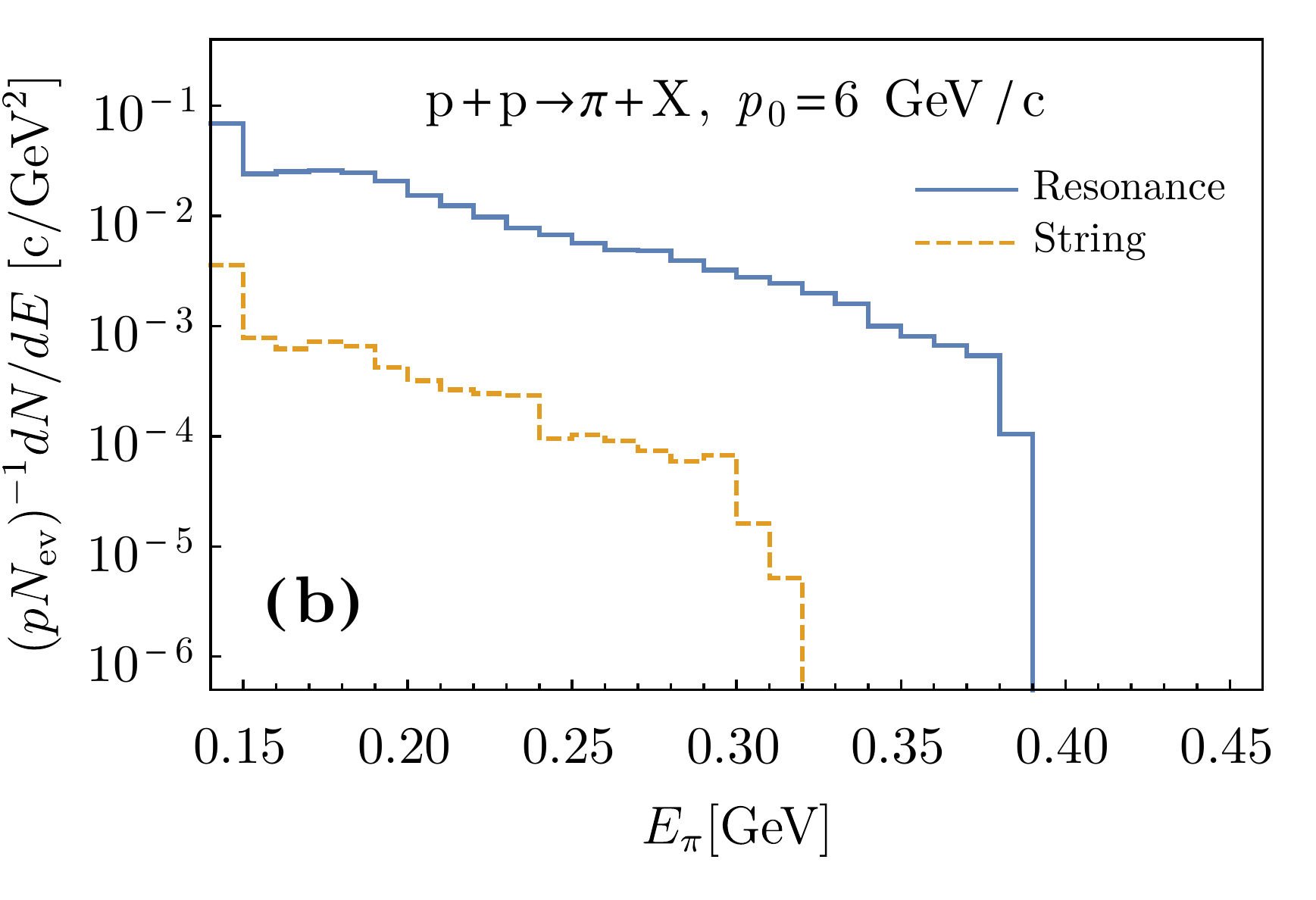}
\includegraphics[width=0.48\textwidth]{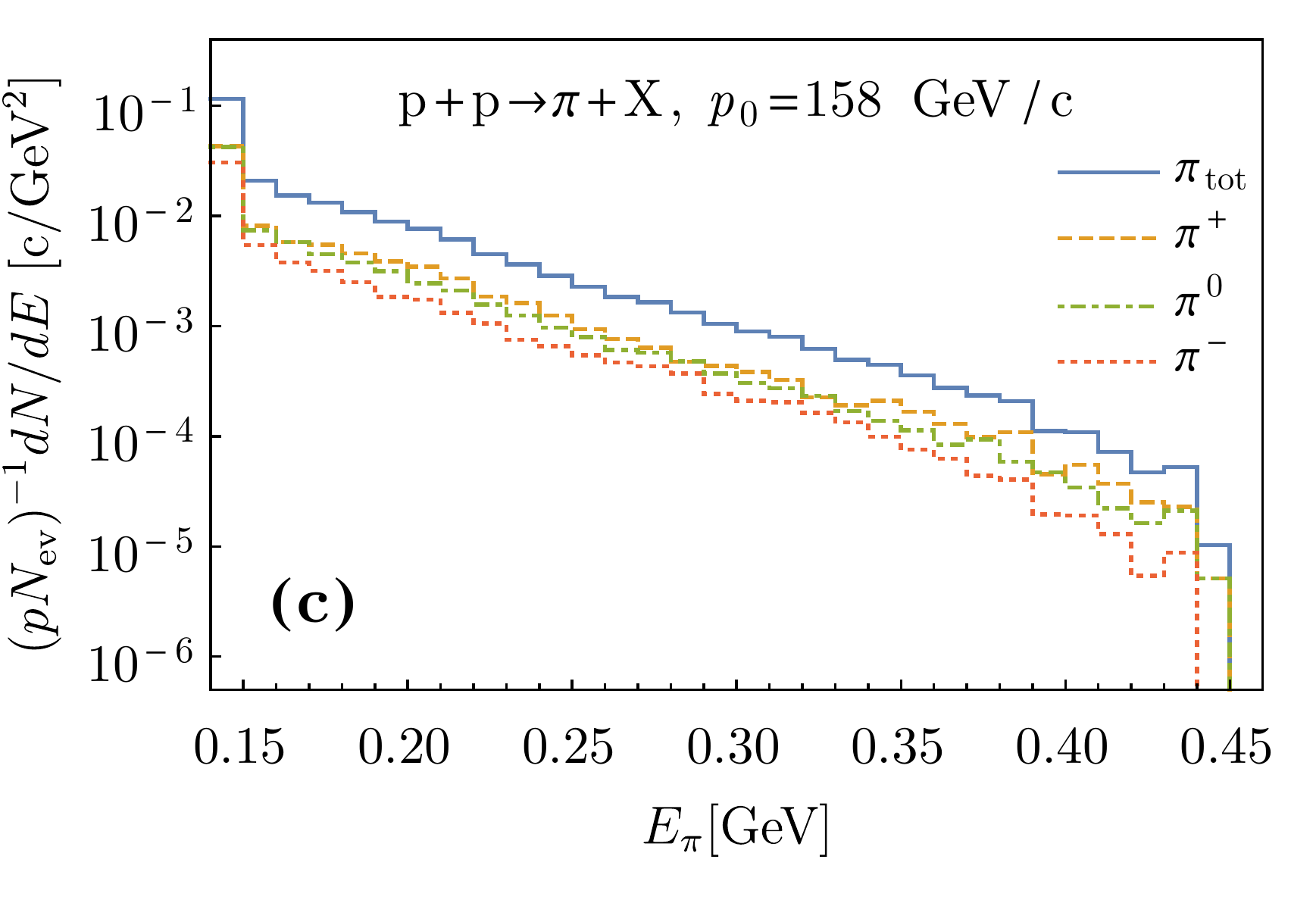}
\includegraphics[width=0.48\textwidth]{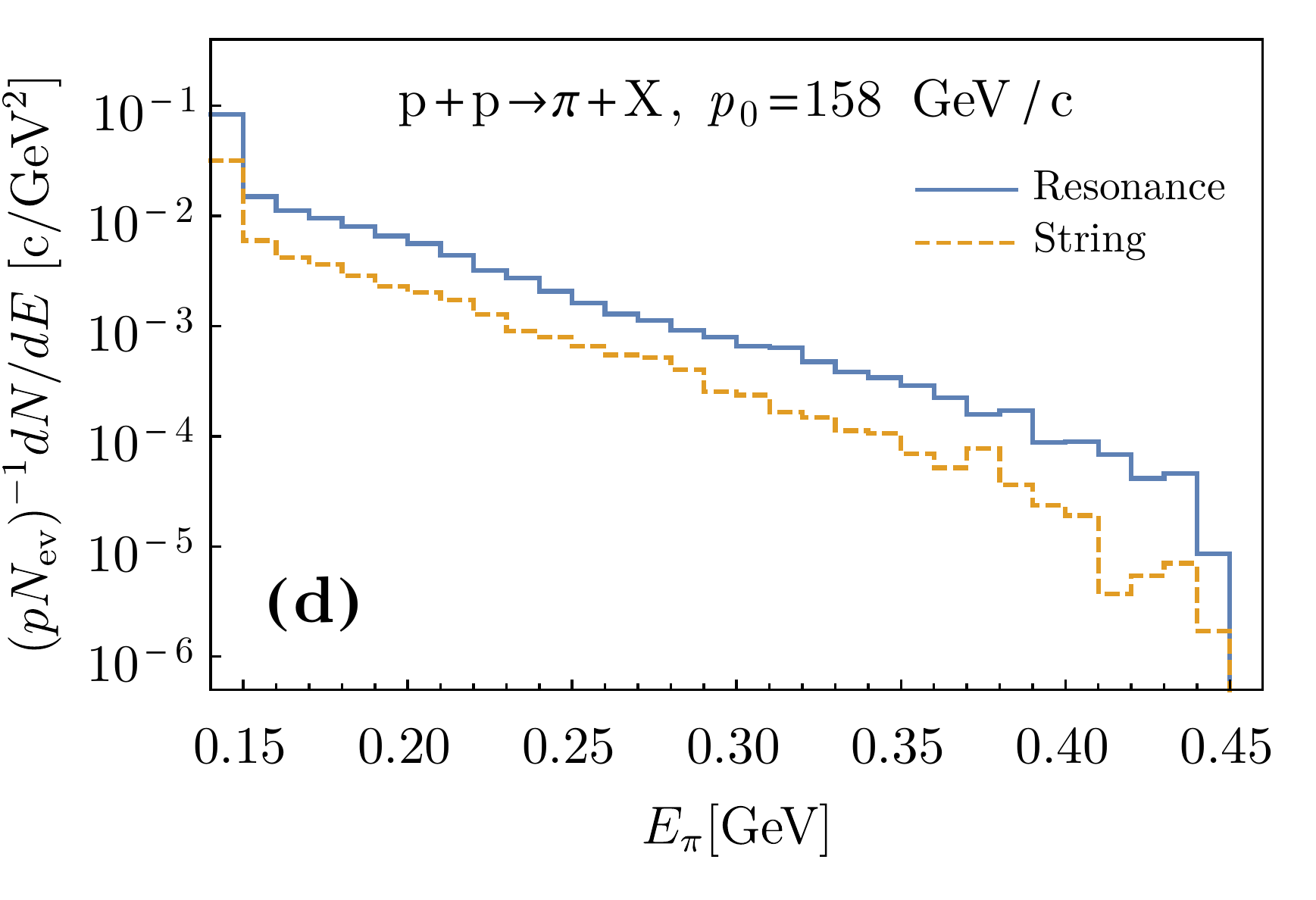}
%\end{floatrow}
\caption{UrQMD results for the pion energy spectra at $180^{\circ}$ in p+p collisions.
 (a): The spectra of $\pi^+$, $\pi^{0}$,
and $\pi^-$ at $p_0=6$ GeV/c. The number of events $N_{\rm ev}=7\cdot 10^7$.
(b): The solid line presents the total pion spectrum from resonance
decays, the dashed line -- from the decay of strings. $p_0=6$~GeV/c, $N_{\rm ev}=7\cdot 10^7$.
(c): Same as in (a) but at $p_0=158$~GeV/c, $N_{\rm ev}=1.4\cdot 10^8$.
%, $N_{\rm ev}=7\cdot 10^7$.
(d): Same as in (b)
but at $p_0=158$~GeV/c, $N_{\rm ev}=1.4\cdot 10^8$. } \label{fig-pp6}	
\end{figure}

First, we make the UrQMD analysis for p+p$\rightarrow\pi(180^\circ)$+X reactions at $p_0=$6~GeV/c and 158~GeV/c.
%For $\pi(180^\circ)$ we choose pions with $p_T/|p_z|<0.1$ which corresponds to $\theta<6^0$.
In p+p and p+A reactions we define the backward pions as those with $\theta=180^\circ\pm6^\circ$ in 
the target rest frame.
In Fig.~\ref{fig-pp6} the energy spectra of pions in reactions p+p$\rightarrow \pi(180^{\circ})+X$
are presented  at $p_0=6$~GeV/c (a) and (b) and  at $p_0=158$~GeV/c (c) and (d).
\begin{figure}[h!]
\centering
%\begin{floatrow}
\includegraphics[width=0.32\textwidth]{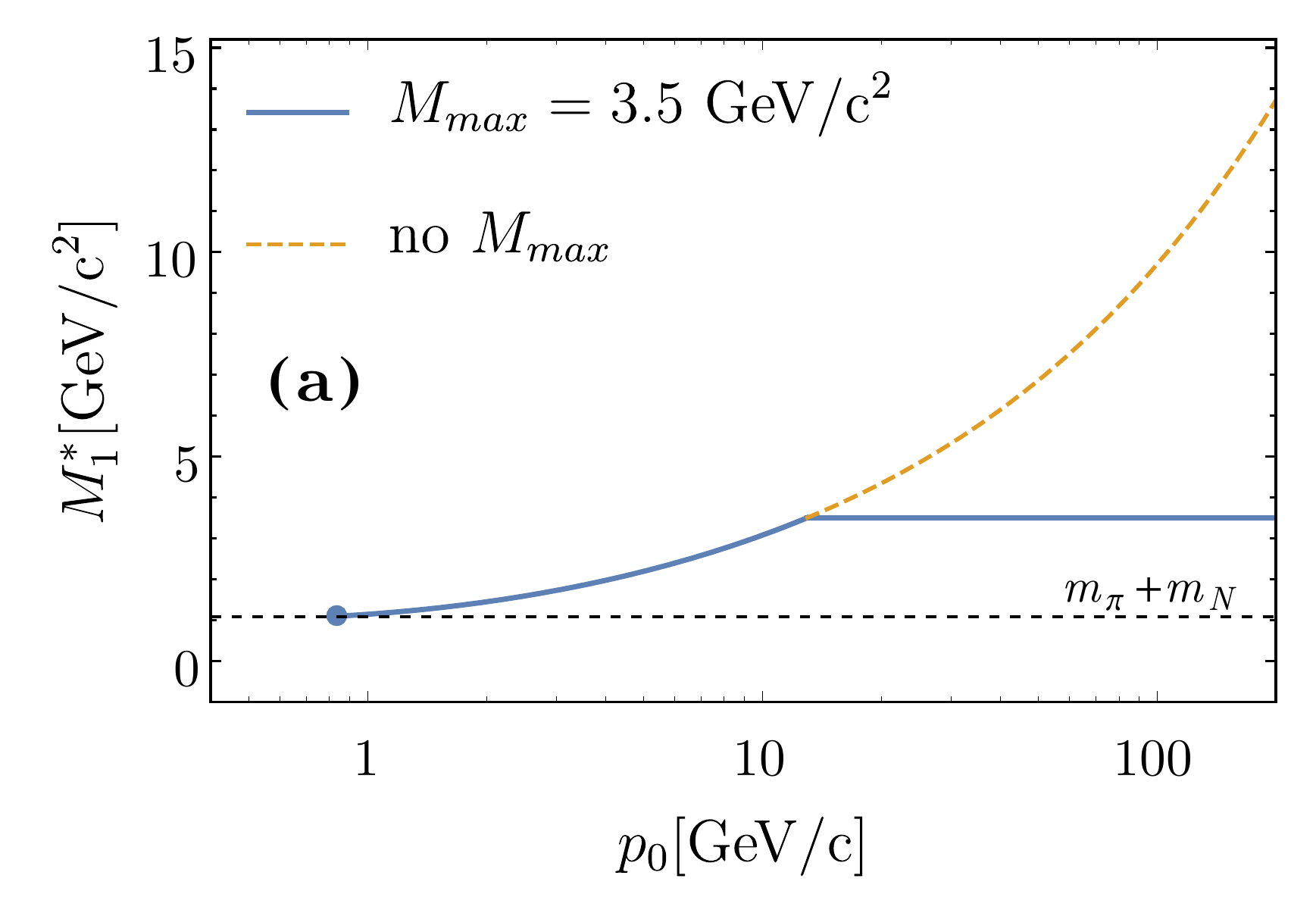}
\includegraphics[width=0.32\textwidth]{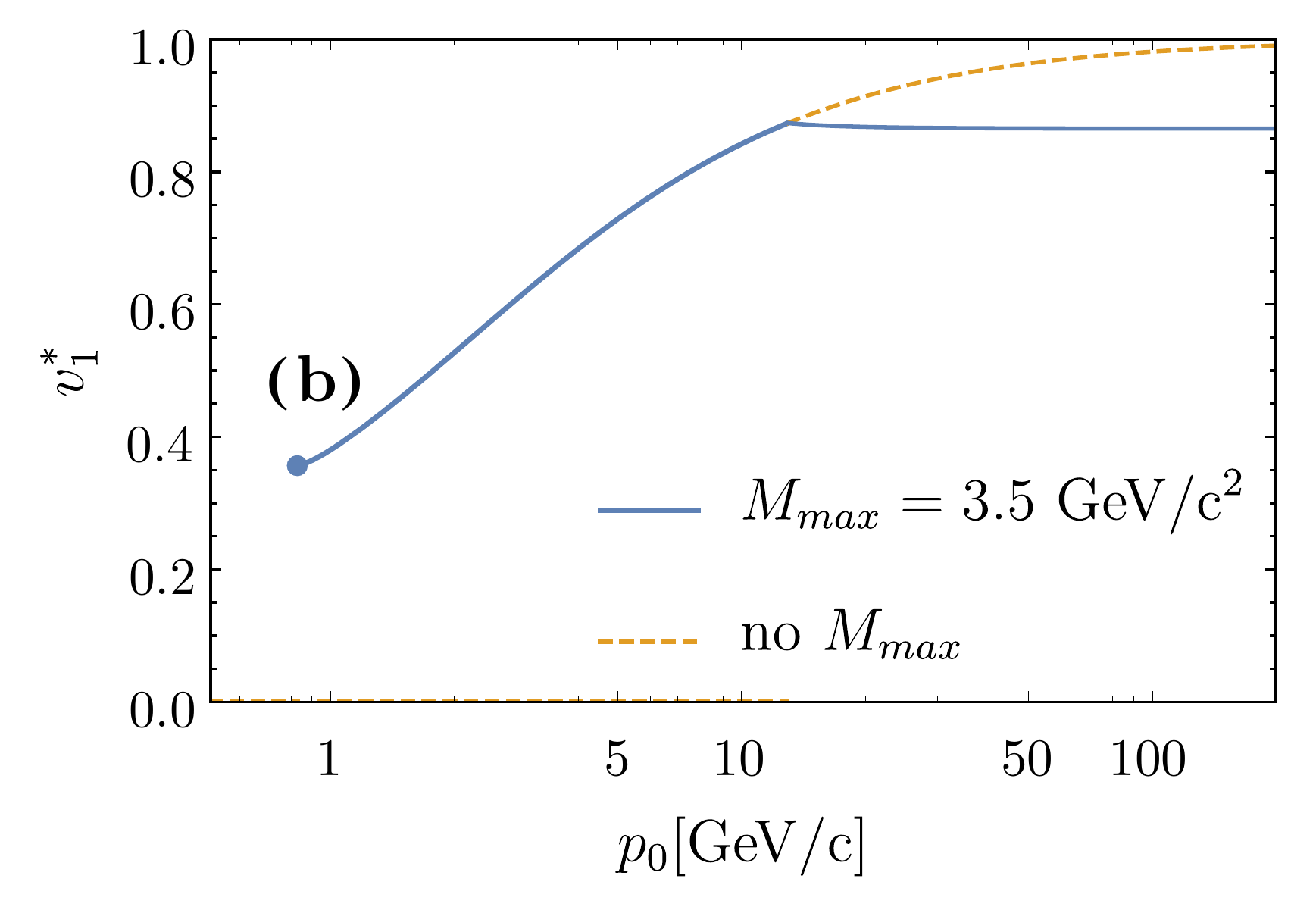}
\includegraphics[width=0.32\textwidth]{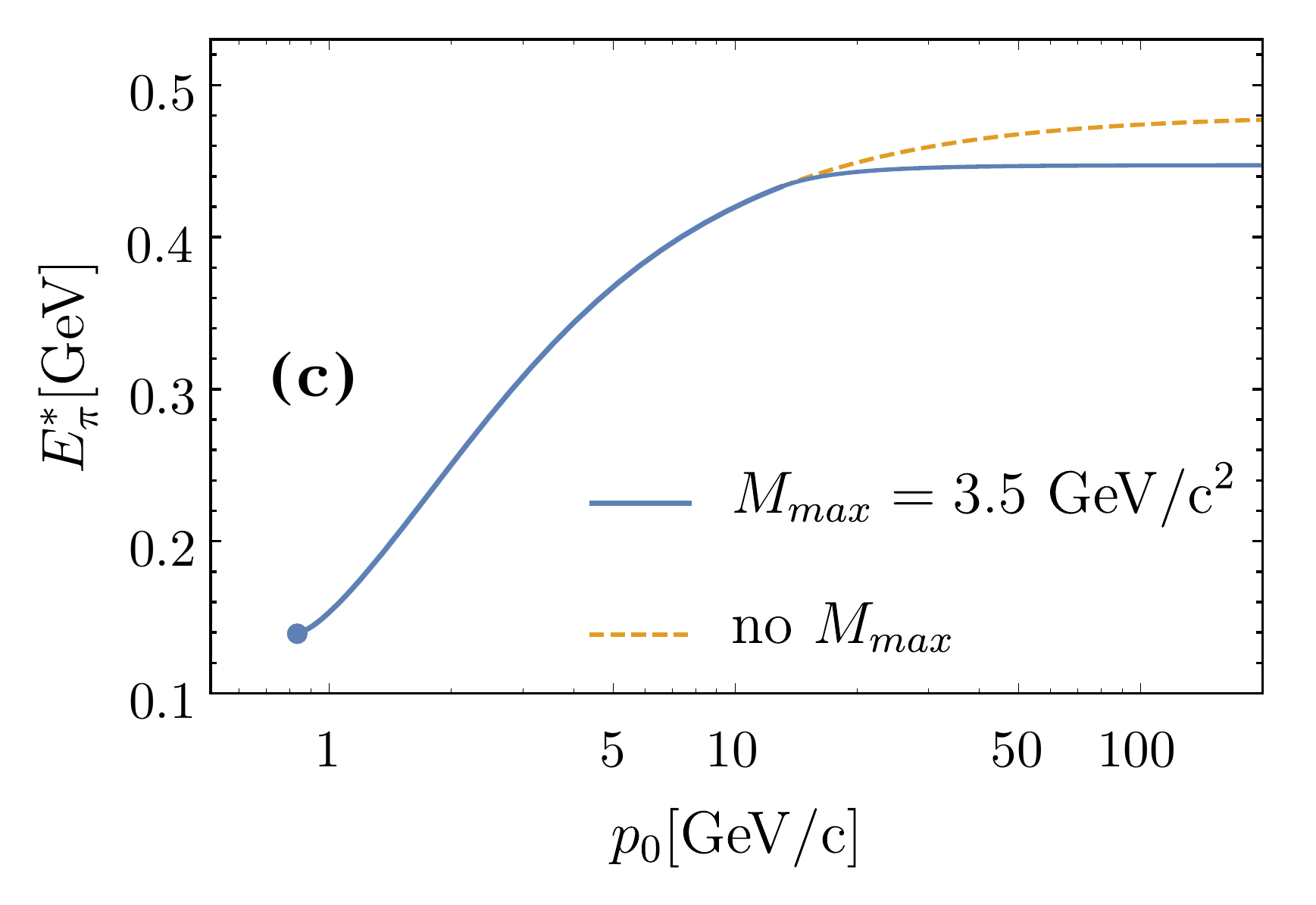}
%\end{floatrow}
\caption{Mass of intermediate resonance $M_R$ (a), its longitudinal velocity
$v_R$ (b), and maximal energy $E_{\pi}^{\rm max}$ of $\pi(180^{\circ})$ (c) as functions of $p_0$.
The upper limit for resonance mass is fixed as $M_{\rm max}$=3.5~GeV$\mathrm{/c^2}$. The dashed lines
correspond to $M_1^*$, $v_1^*$, and $E_{\pi}^*$ shown in Figs.~\ref{fig-M} and \ref{fig-NN}.
}	
\label{E-mass-restrict}
\end{figure}
In Fig.~\ref{fig-pp6} (a) and (c) the energy spectra of $\pi^+$, $\pi^{0}$, and $\pi^-$
are presented  at $p_0=6$~GeV/c and $p_0=158$~GeV/c, respectively. As seen from
Fig.~\ref{fig-pp6} (a) the largest energy
$E_\pi$ for $\pi^+$ and $\pi^{0}$ reaches the maximal possible value $E_\pi^*=0.38$~GeV
for $p_0$=6~GeV/c (see Fig.~\ref{fig-NN}).
These pions are produced in reactions p+p$\rightarrow $p+n+$ \pi^+$ and p+p$\rightarrow $p+p+$ \pi^{0}$.
The mass of intermediate baryonic resonance needed for $\pi(180^{\circ})$ production with energy
$E_\pi^*=0.38$~GeV is $M_R\cong $2.4~GeV$\mathrm{/c^2}$ (see Fig.~\ref{fig-M} (a)).
The number of produced $\pi^-$ in the backward direction
is smaller than that of $\pi^+$.  Besides, the maximal value
of $E_{\pi^-}$ is noticeably smaller than $E_\pi^*=0.38$~GeV.
This is because one needs to produce at least one extra $\pi^+$, i.e.,
p+p$\rightarrow $p+p+$\pi^++\pi^-$,  to satisfy the electric charge conservation.
Figure~\ref{fig-pp6} (b) shows the contributions from
resonances (solid line) and strings (dashed line) to the
total pion energy spectrum at $180^{\circ}$,
i.e., to the sum over all isospin states shown in Fig.~\ref{fig-pp6} (a).
The contribution from string is essentially smaller than that from resonances,
and the maximal possible pion energy $E_\pi^*=0.38$~GeV is not reached.
This is because the two baryonic strings are excited in a collision of two baryons.
As a consequence, at least {\it two} pions should be emitted in the decay of both strings.

Figures \ref{fig-pp6} (c) and (d) present the UrQMD results in p+p reactions at $p_0$=158~GeV/c.
The difference between $\pi^+$ and $\pi^-$ spectra becomes smaller,
and the maximal pion energy  $E_\pi^{\rm max}\cong 0.45$~GeV is approximately
the same. This $E_\pi^{\rm max}$ energy is smaller than $E_\pi^*=0.48$~GeV seen
in  Fig.~\ref{fig-NN}.
To reach the pion energy $E_\pi^*=0.48$~GeV the mass of intermediate baryonic
resonance needed should be
$M_R\cong $12.2~GeV$\mathrm{/c^2}$ (see Fig.~\ref{fig-M} (a)).
However, in the UrQMD there are no resonances with such large masses.

Let us find the value of maximum energy for $\pi(180^{\circ})$ when there is
an upper limit for intermediate resonance mass, i.e., $M_{R}\le M_{\rm max}$.
The energy-momentum conservation in this case is similar
to Eq.~(\ref{pN-cons}), but with additional constraint that $M_1^*$
in Eq.~(\ref{M1*}) should satisfy inequality $M_1^*\le M_{\rm max}$. 
For illustrative purposes we choose $M_{\rm max}=3.5$~GeV$\mathrm{/c^2}$.
The values of $M_R$, $v_R$, and $E_\pi^{\rm max}$ as functions of projectile momentum
$p_0$ are shown in
Fig. \ref{E-mass-restrict}. Almost constant values of $v_R$ and $E_\pi^{\rm max}$
are observed at large $p_0$. In fact, $v_R$ is slightly increasing and $E_\pi^{\rm max}$ slightly
decreasing, but this is hardly seen in Figs. \ref{E-mass-restrict} (b) and (c), respectively.

\subsection{Comparison with the data in p+A collisions}
The UrQMD simulation of p+A reactions were done for most central collisions
with zero impact parameter. To compare our results for the pion spectra with available
data for the inclusive cross sections in p+A collisions we introduce additional normalization factor
considered as a free parameter.
% As it's hard to find correct normalization factor for cross section from computer simulation we just multiplied our cross section with %factor that our cross section became equal to experimental at point $E_\pi=0.25 GeV$. So now we can compare experimental data and data %from UrQMD:

\begin{figure}[h!]
\centering
\includegraphics[width=0.48\textwidth]{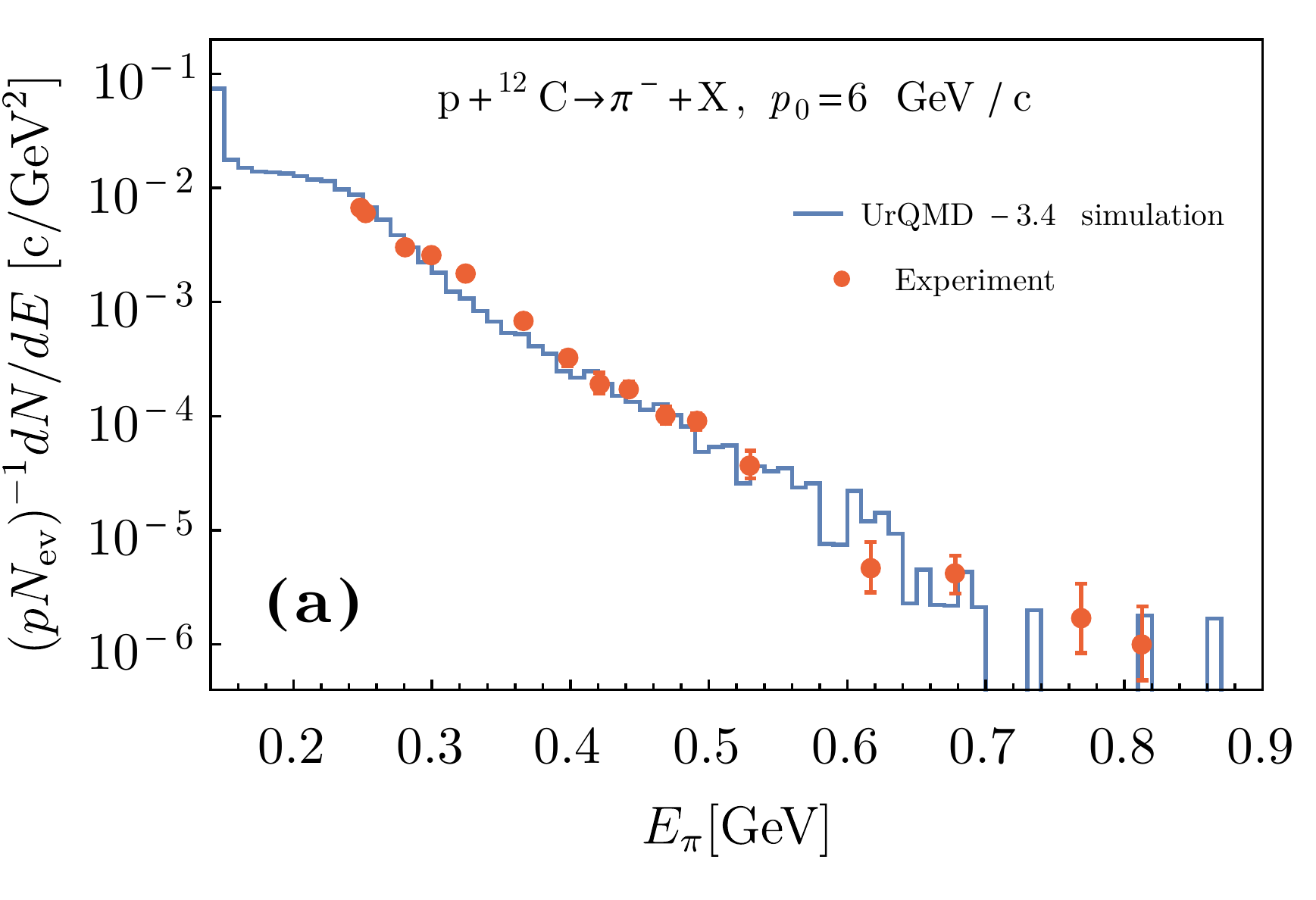}
\includegraphics[width=0.48\textwidth]{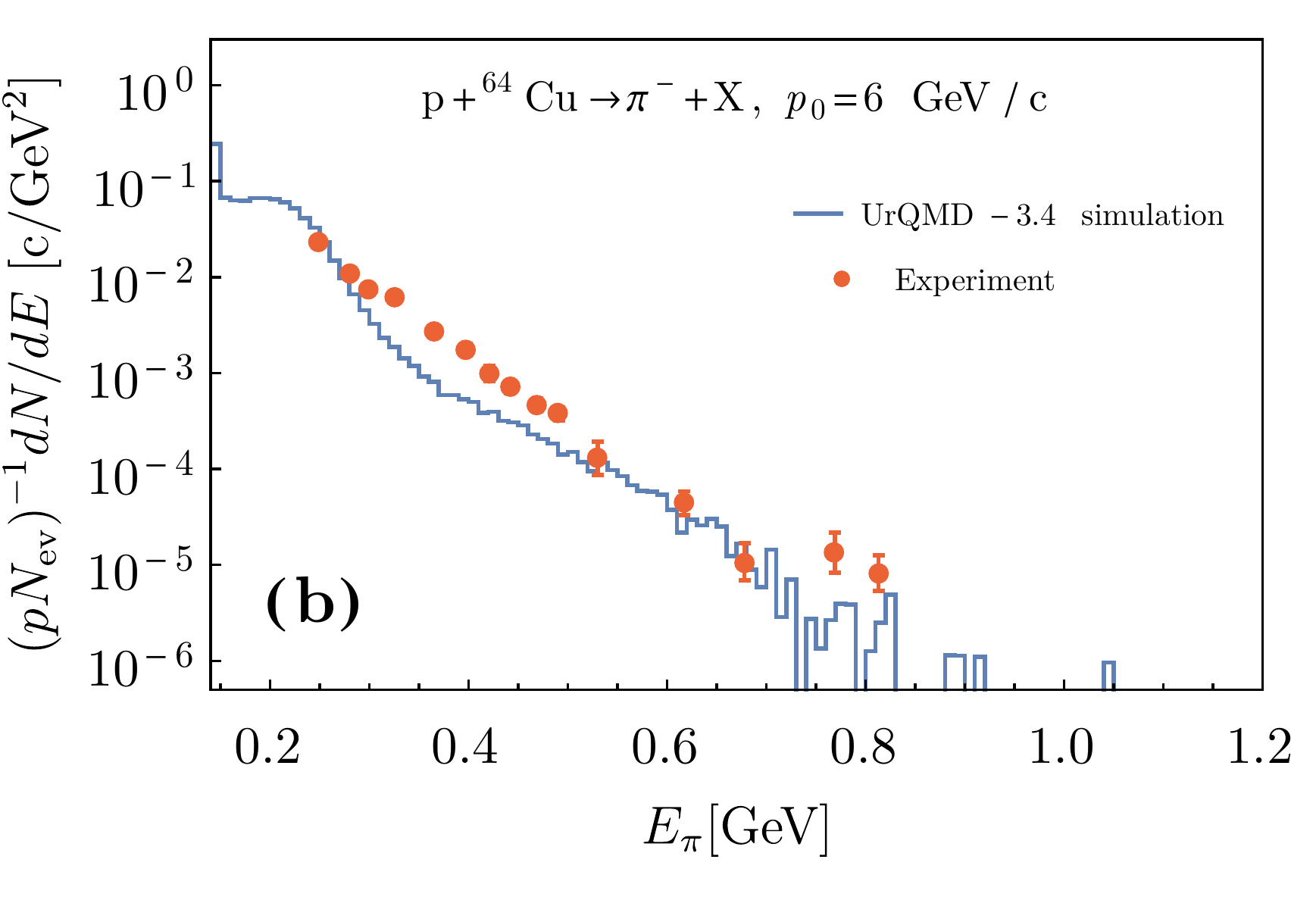}
    \caption{Comparison of the UrQMD results for $\pi ^-$ spectra at 180$^{\circ}$
  with existing data\cite{baldin-74} in p+C and p+Cu reactions  at $p_0$=6~GeV/c, $N_{\rm ev}=10^8$.}\label{fig-data}
\end{figure}
In Fig.~\ref{fig-data} a comparison of the UrQMD results with the data \cite{baldin-74}
is shown for reactions p+C$\rightarrow \pi^- +X$ (a) and p+Cu$\rightarrow \pi^- +X$ (b)  at $p_0=6$~GeV/c.
One may conclude that UrQMD gives a satisfactory  description of the shape of energy spectra
for $\pi^-(180^{\circ})$ at this energy.

More data on cumulative pion production
in p+A and A+A reactions at $p_0$=4.5~GeV/c are presented in Ref.~\cite{bond},
and their analyzes within the FRITIOF model
in Refs.~\cite{Fritiof} and \cite{Fritiof-1}.

\subsection{UrQMD analysis of p+A reactions}
\begin{figure}[h!]
\centering
\includegraphics[width=0.48\textwidth]{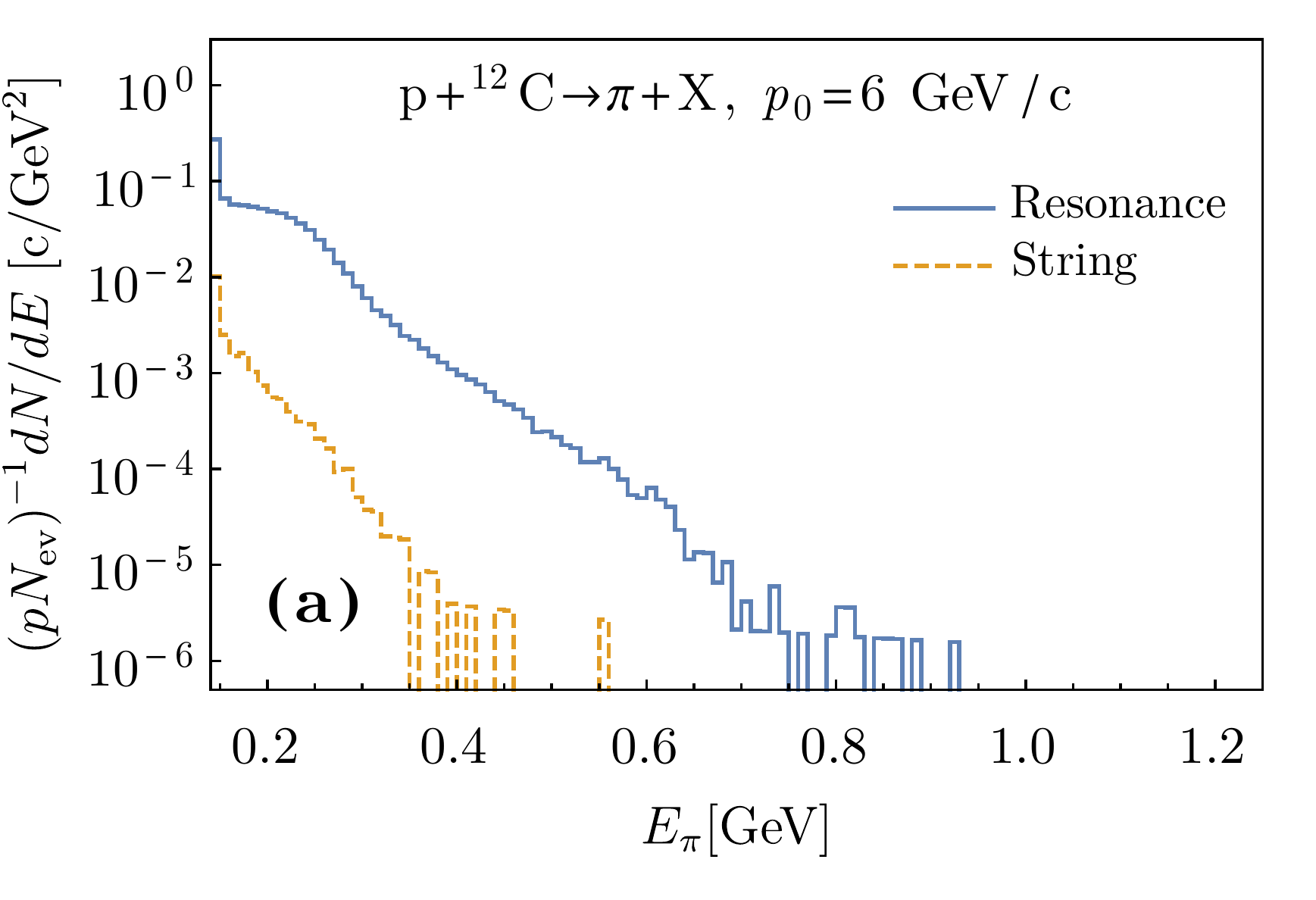}
\includegraphics[width=0.48\textwidth]{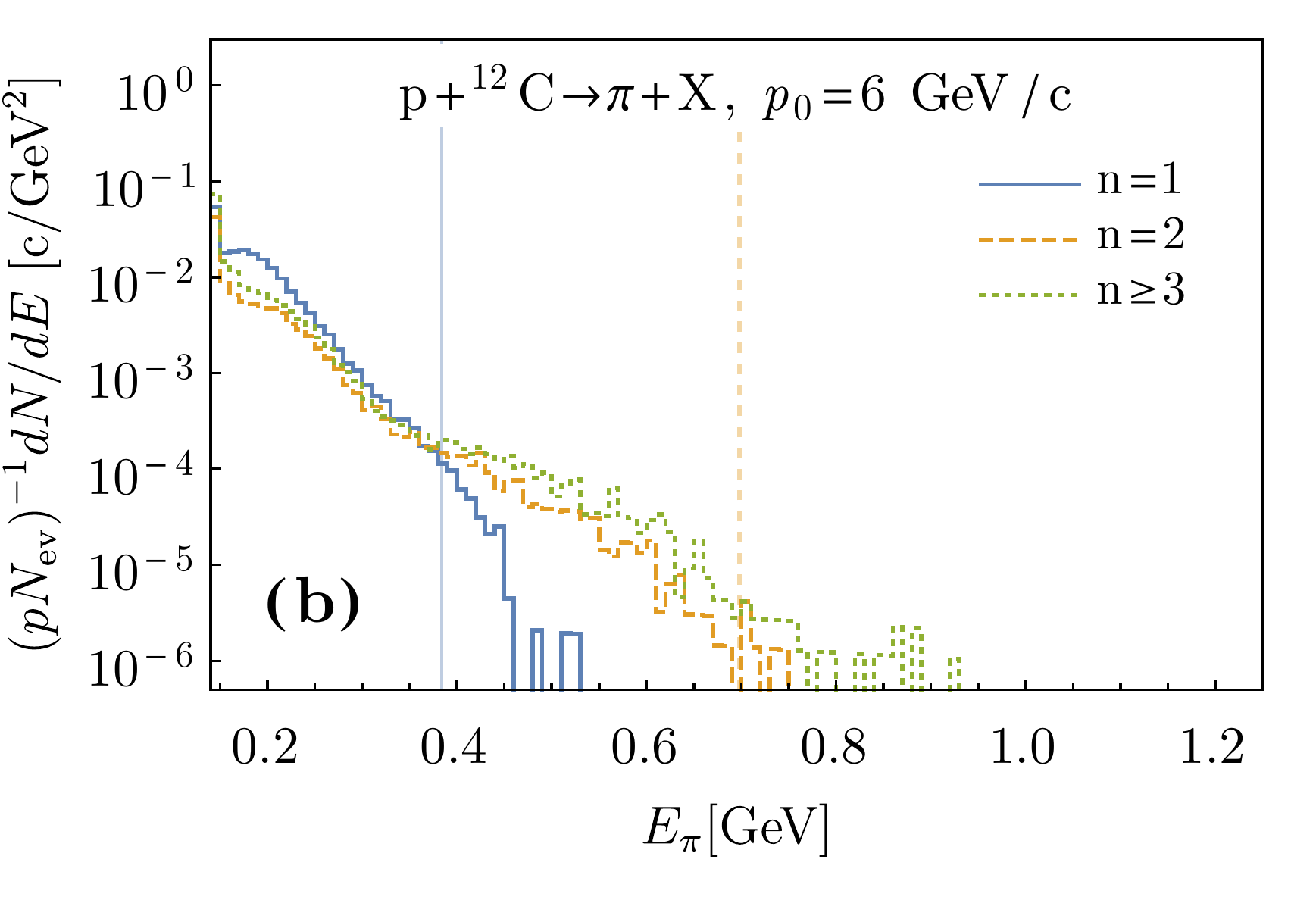}
\includegraphics[width=0.48\textwidth]{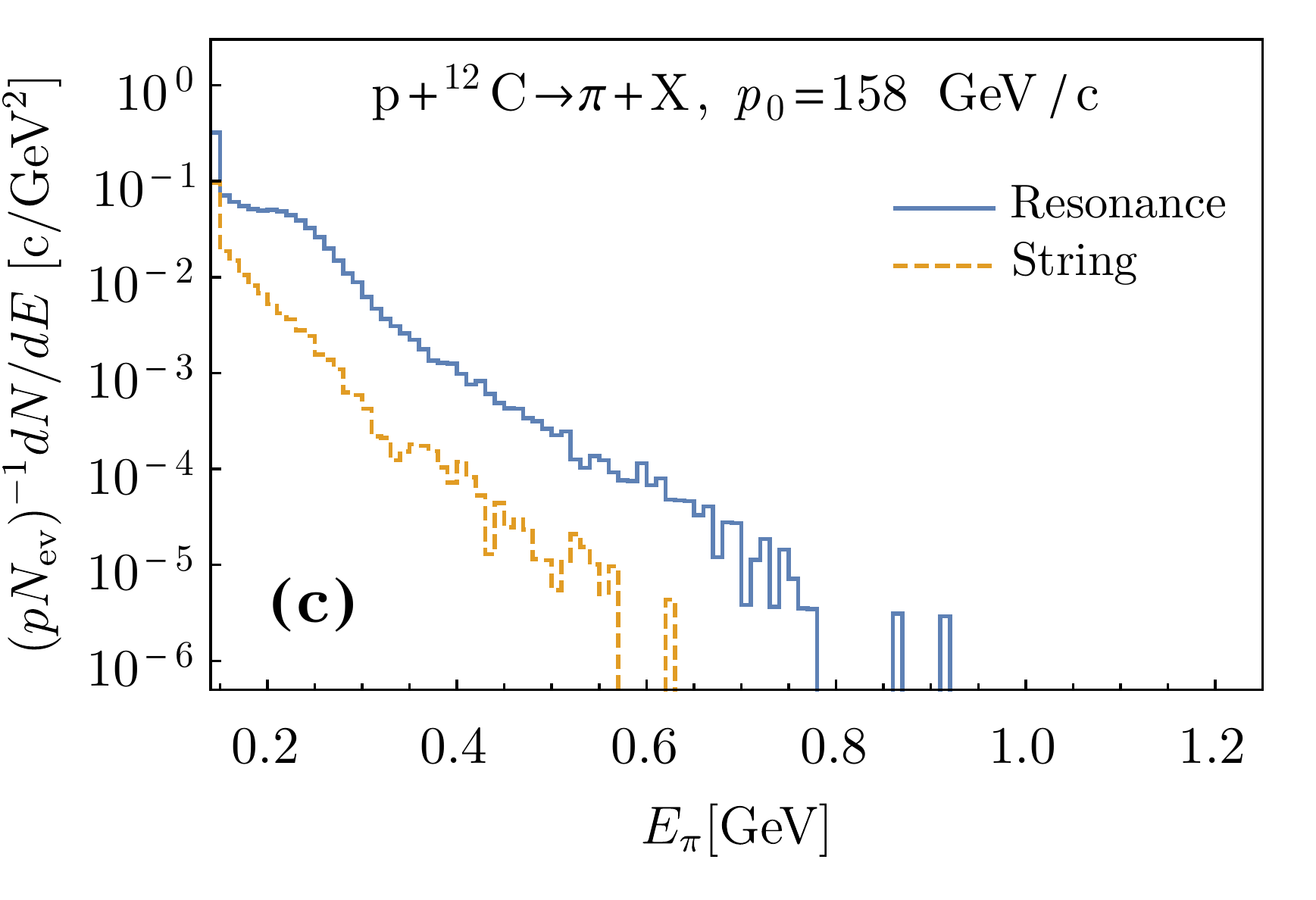}
\includegraphics[width=0.48\textwidth]{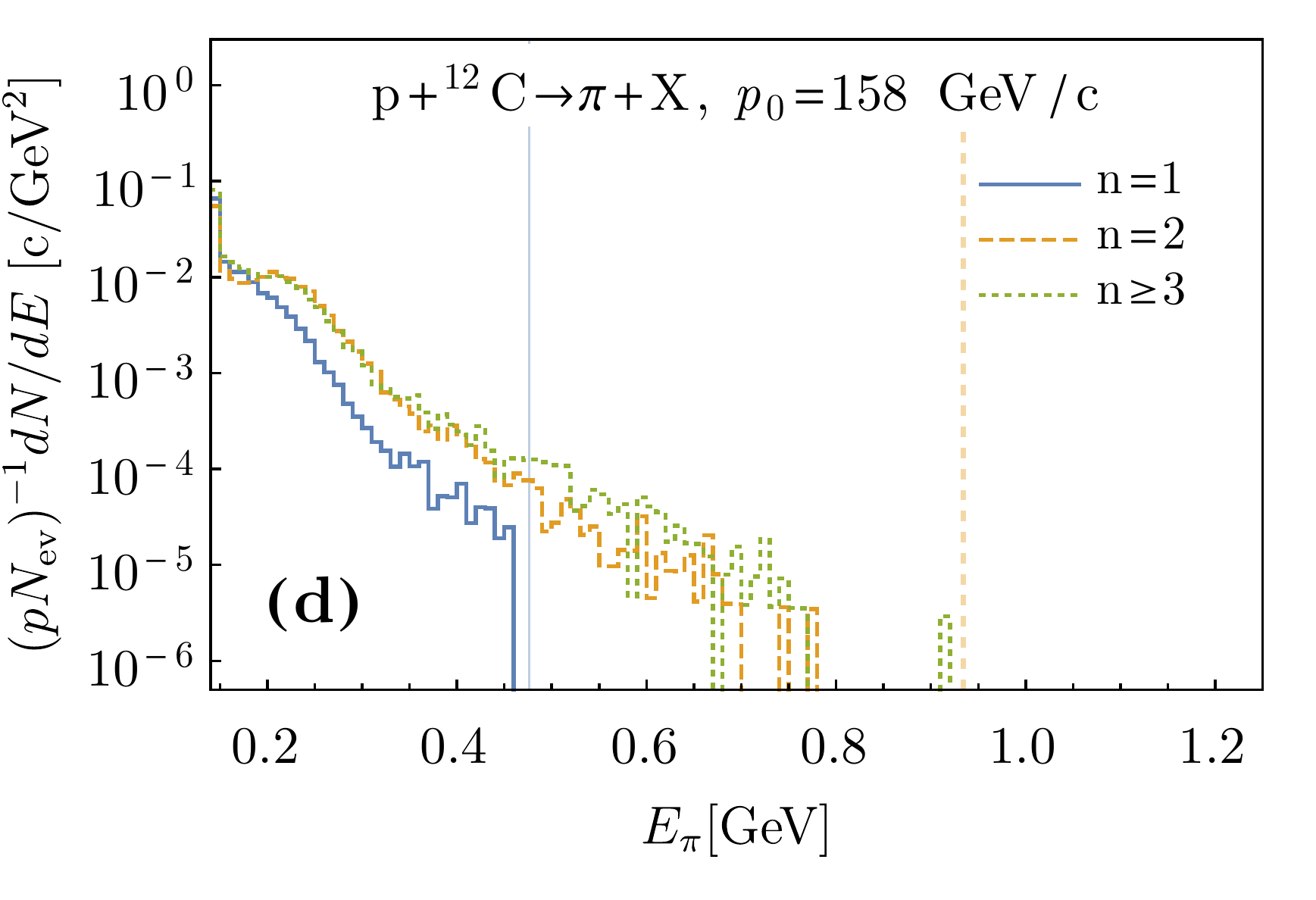}
\caption{UrQMD results for the pion energy spectra at $180^{\circ}$ in p+$^{12}$C collisions.
 (a): The solid line presents the total pion spectrum from resonance
decays, the dashed line -- from the decay of strings.
(b): Spectra for pions created after different number of collisions with nuclear nucleons:
$n=1$ (solid line) collision, $n=2$ (dashed line), and $n\ge 3$ (dotted line).
Vertical lines correspond to energies $E^*_{\pi,1}$ (solid) and $E^*_{\pi,2}$ (dashed).
In (a) and (b), $p_0=6$~GeV/c and the number of events  $N_{\rm ev}=10^9$.
(c) and (d): Same as in (a) and (b), respectively, but at $p_0=158$~GeV/c with
$N_{\rm ev}=3.8\cdot 10^7$.
%created in $n$ successive collisions(dashed -- $E^*_\pi$ for $p_0=6$ GeV/c -- (a),
%and created in resonance or string decay -- (b)
}	
\label{fig-pC}
\end{figure}

In this subsection we present the UrQMD simulations
for central p+A$\rightarrow \pi(180^{\circ})+$X collisions and make their microscopic analysis.
The total pion spectra summed over all charges will be presented.
The UrQMD gives a good opportunity
to study a history of each individual reaction. Let us start from the results for
p+C reactions at $p_0=6$~GeV/c. In Figs.~\ref{fig-pC} (a) the
contributions to $\pi(180^{\circ})$ spectrum from the decays of resonances (solid line)
and strings (dashed line) are presented.
\begin{figure}[h!]
\centering
\includegraphics[width=0.48\textwidth]{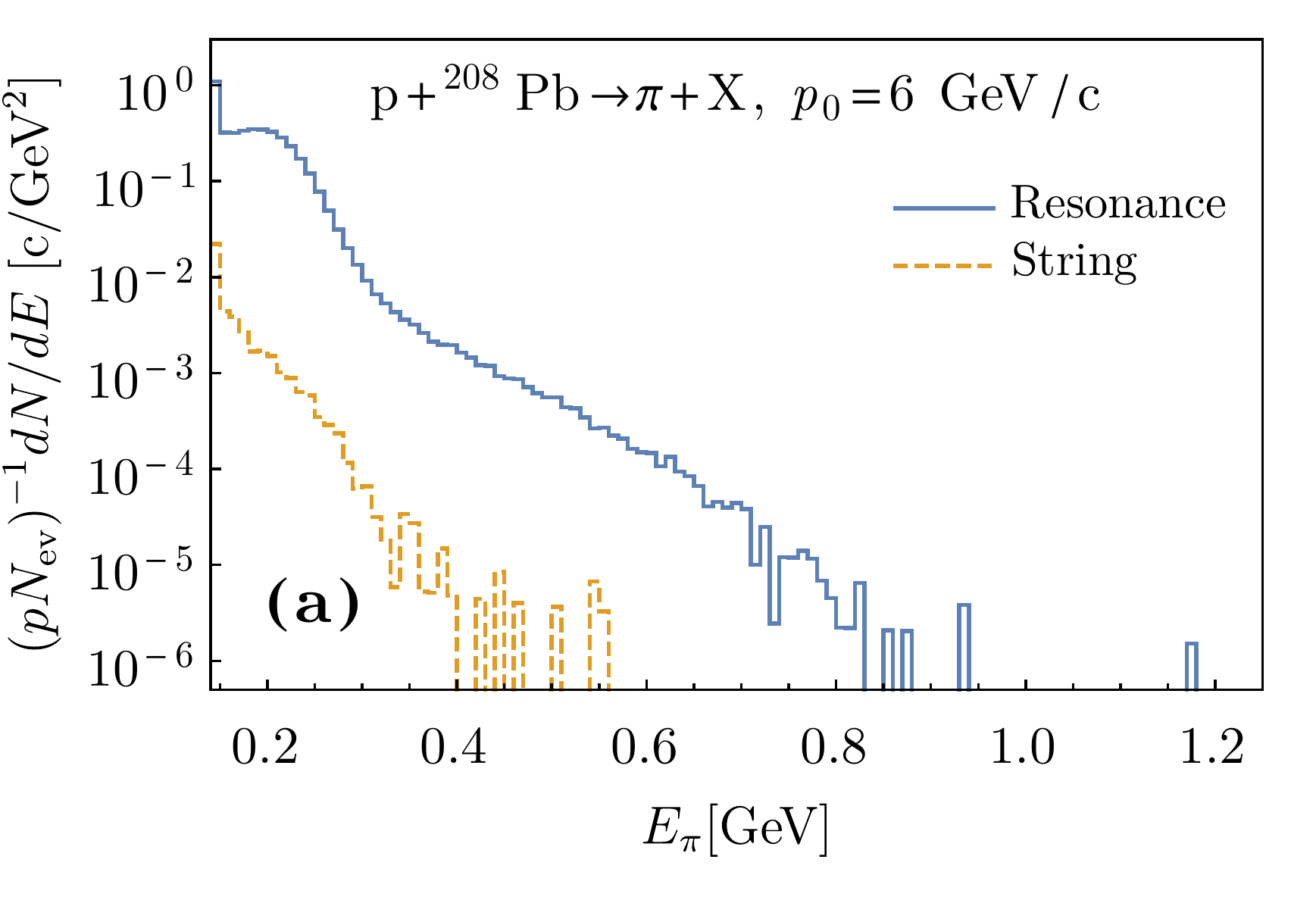}
\includegraphics[width=0.48\textwidth]{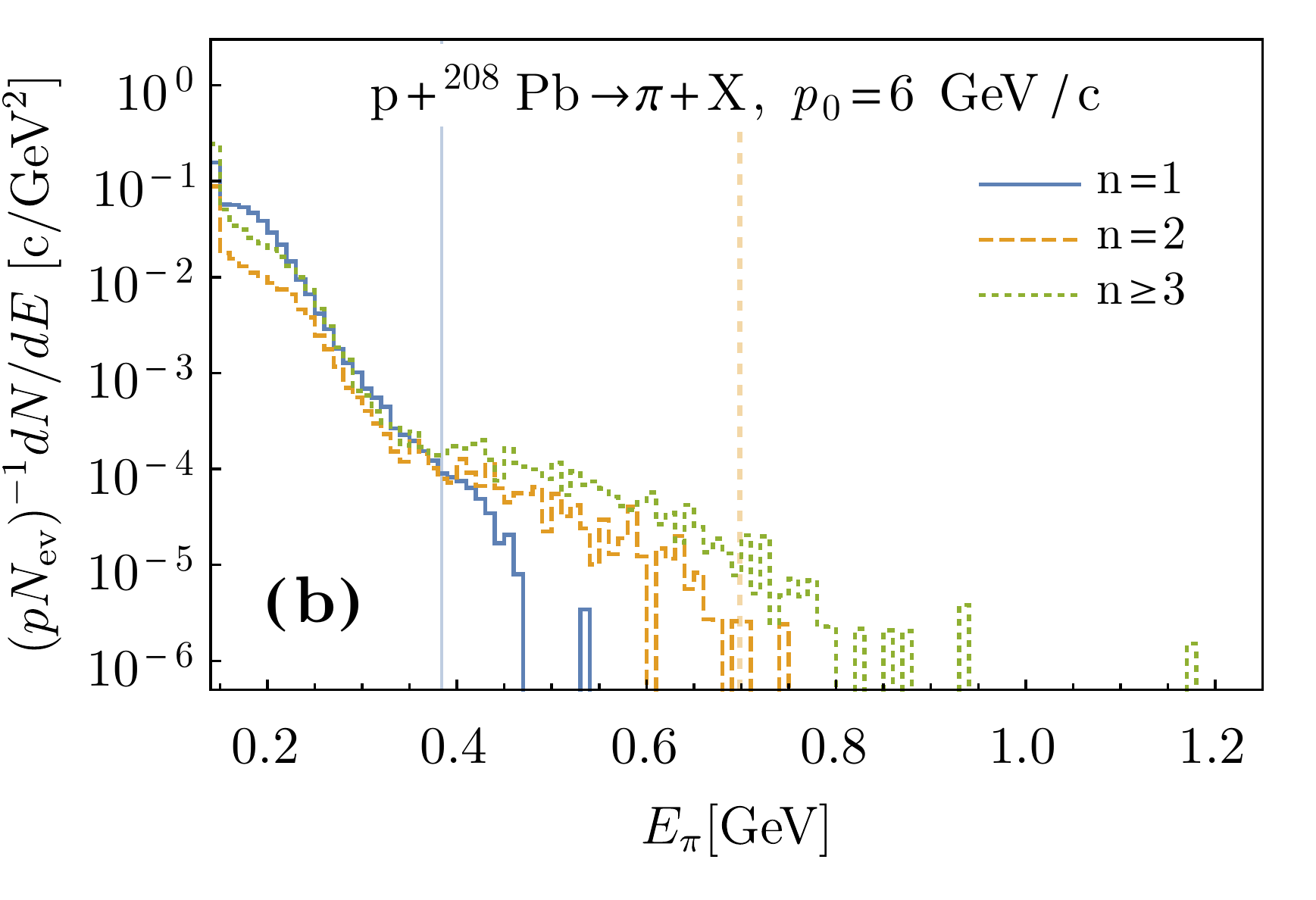}
\includegraphics[width=0.48\textwidth]{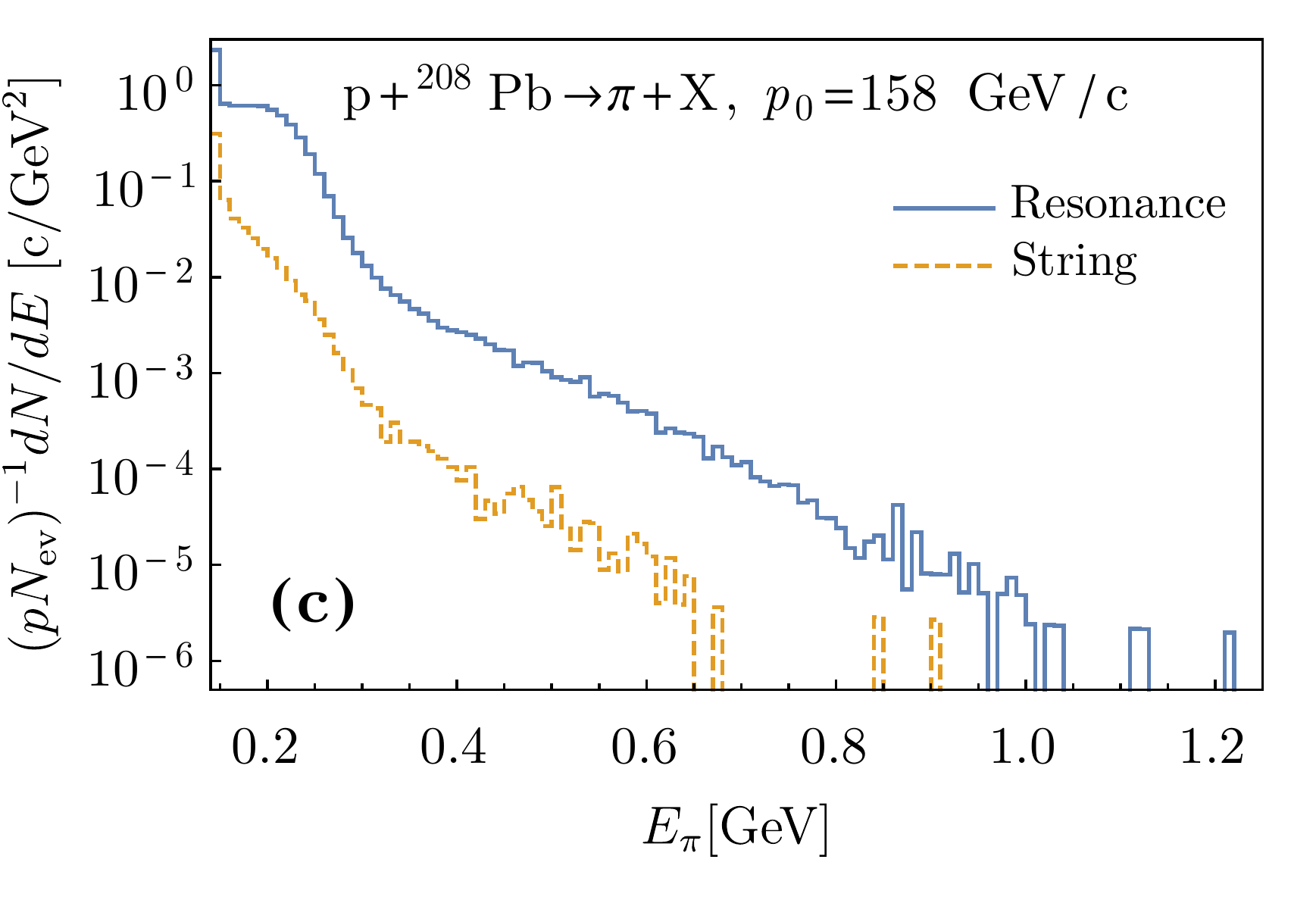}
\includegraphics[width=0.48\textwidth]{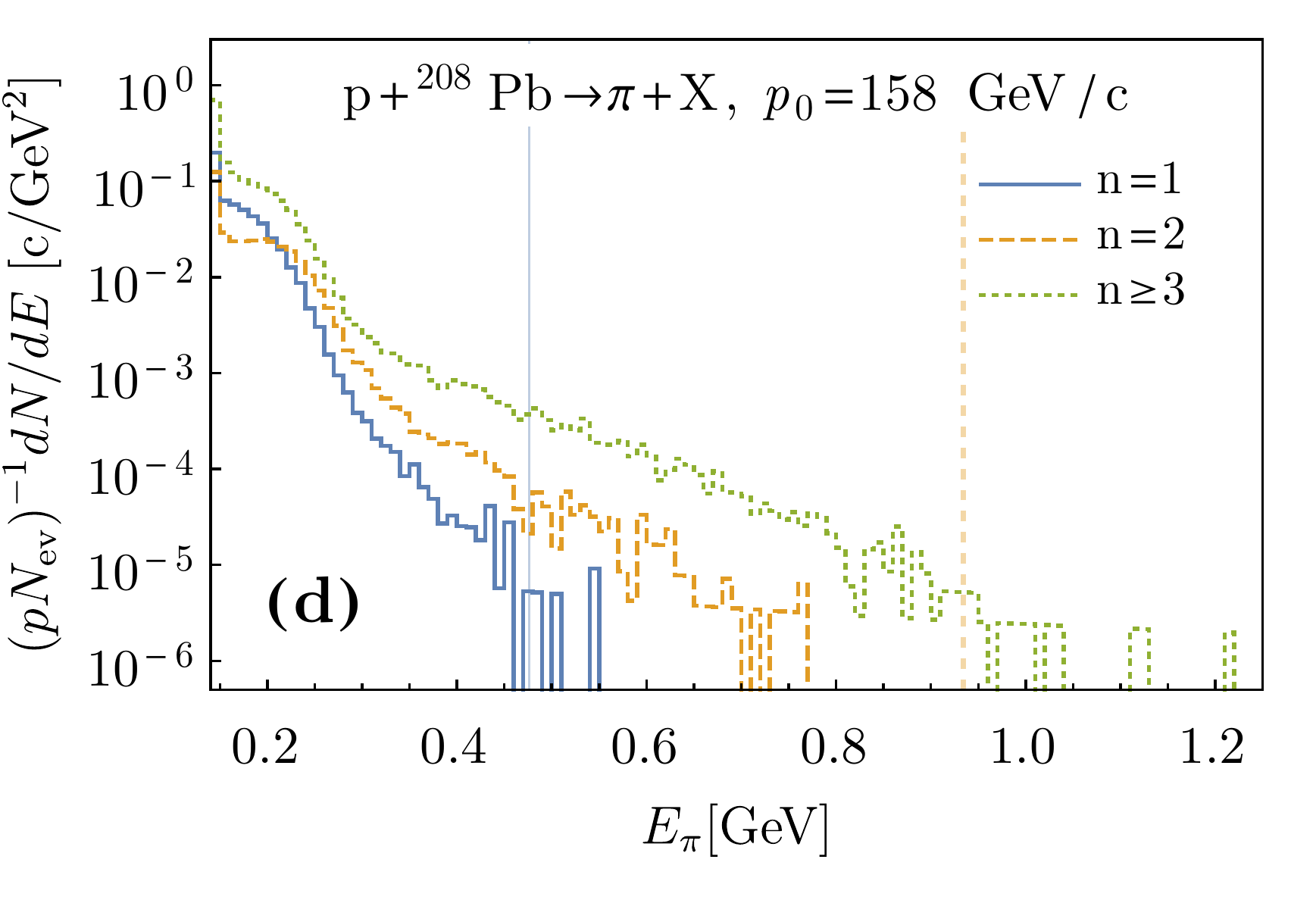}
\caption{Same as in Fig.~\ref{fig-pC} but for p+$^{208}$Pb collisions.
 (a) and (b):
 %The solid line presents the total pion spectrum from resonance
%decays, the dashed line -- from the decay of strings. The number of events
$N_{\rm ev}=7\cdot 10^7$.
%(b):
%Spectra for pions that were created after different number of collisions with nuclear nucleons, solid line -- in one collision, dashed -- %after two successive collisions, dotted -- three or more collisions. Solid vertical line corresponds to maximal energy at this projectile %momentum in one collision -- $E^*_{\pi,1}$, dashed line -- in two successive collisions $E^*_{\pi,1}$. $p_0=6$~GeV/c,
%$N_{\rm ev}=5.7\cdot 10^7$.
(c) and (d):
%Same as in (a) but at $p_0=158$~GeV/c.
%, $N_{\rm ev}=7\cdot 10^7$.
%(d): Same as in (b)
%but at $p_0=158$~GeV/c,
$N_{\rm ev}=4.2\cdot 10^7$.}
%Spectra for $\pi(180^\circ)$ which were created in $n$ successive collisions(dashed -- $E^*_\pi$ for $p_0=6$ GeV/c -- (a), spectra for %$\pi(180^\circ)$ that were created in resonance or string decay -- (b)
%
\label{fig-pPb}
\end{figure}
%%%%%%

%
The resonance contribution is evidently a dominant one. The next question concerns
the history of each resonance contributing to the $\pi(180^{\circ})$ spectrum. Particularly,
how many successive collisions with nuclear nucleons take place?
%Using detailed history of collisions we made an analysis of number of collisions $n$ of every $\pi(180^\circ)$.
%To study number of collisions we selected events where projectile proton participated in reactions $2\rightarrow X$.
In Fig.~\ref{fig-pC} (b) we compare the spectra of $\pi(180^{\circ})$ emitted from resonance
decay after $n=1$ (solid line), $n=2$ (dashed line),
and $n\geq 3$ (dotted line) successive collisions of the projectile with nuclear nucleons.
The vertical solid and dashed lines
show the values of $E^*_{\pi,1}\cong 0.38$~GeV and $E^*_{\pi,2}\cong0.70$~GeV  in Fig.~\ref{fig-pC} (b)
and $E^*_{\pi,1}\cong 0.48$~GeV and $E^*_{\pi,2}\cong0.93$~GeV  in Fig.~\ref{fig-pC} (d).
These quantities
%$E^*_{\pi,1}$ and $E^*_{\pi,2}$ presented in Fig.~\ref{fig-En}
correspond to the kinematical limits
for the pion energy emitted at $180^{\circ}$ for, respectively, one and two successive
collisions with nuclear nucleons
at given value of initial momentum $p_0$.

From Figs.~~\ref{fig-pC} (b) and (d) one observes that $E_\pi$ may exceed $E^*_{\pi,1}$ even
for $n=1$ contribution. This happens because of nucleon motion inside nuclei (Fermi motion)
which exists in the UrQMD model. This effect is, however, not large. The main contribution
to the kinematical region forbidden for p+N collisions (i.e., to $E_\pi > E^*_{\pi,1}$)
comes from the decays of resonances created within $n=2$ and $n\ge 3$ successive collisions with nuclear nucleons.
These are the massive baryonic resonances that are included in the UrQMD model.
Note that the role of heavy baryonic resonances within the UrQMD simulations was also discussed
in Refs.~\cite{Bleicher,Bleicher-1,Vovchenko}.

\section{Summary}\label{sec-sum}
Pion production in the backward direction in
the target rest frame in p+A collisions is considered .
Pions created  outside the kinematical boundary of p+N reactions
are  studied. This is the so called cumulative effect discovered in p+A
reactions about 40 years ago. We develop the model, where
particle production in the cumulative
region is due to the formation of  massive resonance states,
successive collisions of this resonances with nuclear nucleons, and
finally their decay
with emission of a cumulative particle.
The restrictions that follow from energy-momentum conservation
are studied. In p+A interactions, the resonances created in p+N reactions may have further
inelastic collisions with nuclear nucleons. Due to successive
collisions with nuclear nucleons, the masses of these resonances may increase and
simultaneously their longitudinal velocities decrease. In our model,
these two effects give an explanation of the cumulative pion production.
The simulations of p+A reactions within the UrQMD model support this
physical picture.

It should be noted that
hadron-like systems with very high masses (heavy resonances, Hagedorn fireballs,
quark-gluon bags, baryon and meson strings) are of primary importance for properties
of strongly interacting matter at high temperatures and/or baryonic density.
They may have also decisive influence on the transition between hadron matter and
quark-gluon plasma, and define the type of this phase transition and existence of the
QCD critical point itself. However, reliable experimental information about the
properties of these heavy hadron-like states, and even about their existence, are absent.
We suggest that further experimental studies of cumulative effect in p+A reactions at
high collision energies are probably the best way to search for signatures
of very heavy hadron-like objects
and investigate their properties.
Particularly, these studies can be done by the NA61/SHINE Collaboration \cite{NA61,NA61-1}
at the SPS CERN, where the hadron production in p+p, p+A, and A+A collisions is investigated.
New efforts may be also required to extend existing
relativistic transport models, like UrQMD and HSD,  by adding to their formulations
string+hadron inelastic reactions that are absent in the present codes of these models.

%%%%%%%%%%%%%%%%%%%%%%%
\begin{acknowledgments}
%%%%%%%%%%%%%%%%%%%%%%%
%
We would like to thank
Elena Bratkovskaya,
Marek Ga\'zdzicki, Volodymyr Vovchenko, and Gennady Zinovjev
for fruitful discussions and comments. We are thankful
to the staff of the computer maintenance department in Bogolyubov Institute for Theoretical
Physics, Kiev for the access to the local cluster computer.
The work of  M.I.G. was supported  by the
Program of Fundamental Research of the Department of Physics and
Astronomy of NAS.
\end{acknowledgments}

\end{document}